\documentclass[10pt]{wlscirep}
\usepackage[utf8]{inputenc}
\usepackage[T1]{fontenc}

\usepackage{mathtools}
\makeatletter
\DeclareRobustCommand\widecheck[1]{{\mathpalette\@widecheck{#1}}}
\def\@widecheck#1#2{%
    \setbox\z@\hbox{\m@th$#1#2$}%
    \setbox\tw@\hbox{\m@th$#1%
       \widehat{%
          \vrule\@width\z@\@height\ht\z@
          \vrule\@height\z@\@width\wd\z@}$}%
    \dp\tw@-\ht\z@
    \@tempdima\ht\z@ \advance\@tempdima2\ht\tw@ \divide\@tempdima\thr@@
    \setbox\tw@\hbox{%
       \raise\@tempdima\hbox{\scalebox{1}[-1]{\lower\@tempdima\box
\tw@}}}%
    {\ooalign{\box\tw@ \cr \box\z@}}}
\makeatother

\usepackage{textcomp}
\usepackage{caption}
\usepackage{subcaption}
\usepackage{comment}
\usepackage{multicol}
\usepackage[ruled,linesnumbered]{algorithm2e}

\usepackage{multirow}
\usepackage{color}

\title{Channel Phase Processing in Wireless Networks for Human Activity Recognition}

\author[1,*]{Guillermo~Diaz}
\author[2]{Iker~Sobron}
\author[1]{Iñaki~Eizmendi}
\author[1]{Iratxe~Landa}
\author[3]{Johana~Coyote}
\author[1]{Manuel~Velez}
\affil[1]{Dept. Communications Engineering, University of the Basque Country (UPV/EHU), Bilbao, Spain}
\affil[2]{Dept. of Computer Languages and Systems, University of the Basque Country (UPV/EHU), Bilbao, Spain}
\affil[3]{Dept. of Telecommunications, National Autonomous University of Mexico (UNAM), Mexico City, Mexico}

\affil[*]{guillermo.diaz@ehu.eus}

\begin{abstract}
The phase of the channel state information (CSI) is underutilized as a source of information in wireless sensing due to its sensitivity to synchronization errors of the signal reception. A linear transformation of the phase is commonly applied to correct linear offsets and, in a few cases, some filtering in time or frequency is carried out to smooth the data. This paper presents a novel processing method of the CSI phase to improve the accuracy of human activity recognition (HAR) in indoor environments. This new method, coined Time Smoothing and Frequency Rebuild (TSFR), consists of performing a CSI phase sanitization method to remove phase impairments based on a linear regression transformation method, then a time domain filtering stage with a Savitzy-Golay (SG) filter for denoising purposes and, finally, the phase is rebuilt, eliminating distortions in frequency caused by SG filtering. The TSFR method has been tested on five datasets obtained from experimental measurements, using three different deep learning algorithms, and compared against five other types of CSI phase processing. The results show an accuracy improvement using TSFR in all the cases. Concretely, accuracy performance higher than 90\% in most of the studied scenarios has been achieved with the proposed solution. In few-shot learning strategies, TSFR outperforms the state-of-the-art performance from 35\% to 85\%.
\end{abstract}
\begin{document}

\flushbottom
\maketitle

\thispagestyle{empty}

\section*{Introduction}

Wireless Sensing has been a rapidly growing field of study within the Internet of Things in recent years. It involves measuring wireless channel characteristics using existing wireless networks, such as WiFi networks, to sense environmental changes in the surrounding area of the network.
Human activity recognition (HAR) in indoor environments is one of the main fields of application of wireless sensing.

The pervasive deployment of wireless networks worldwide and the fact that wireless sensing can be considered a privacy-preserving solution make this technology a promising alternative to other sensing solutions such as video surveillance with depth cameras or wearables \cite{kinect2,kinect3,kinect1,kinect4,wereable2, wereable}. Those other sensing methods present some drawbacks; for instance, cameras can compromise user privacy and, in the case of wearables, users should carry the devices on their person to be monitored. In wireless systems based on orthogonal frequency-division multiplexing (OFDM), such as WiFi, the Received Signal Strength Indicator (RSSI) and the Channel State Information (CSI) are used for wireless sensing.  RSSI suffers from significant uncertainties due to the signal fluctuations under actual conditions, such as scattering, degradation, and sensitivity to noise \cite{rssi}. Therefore, in recent years, CSI data have been widely used due to its major robustness against noise and other impairments of the signal reception. A time series of CSI measurements show how wireless signals propagate through objects and humans in the time, frequency, and spatial domains and can be used for different monitoring applications. Due to this, human activity recognition is an important field of wireless sensing, ranging in several areas such as crowd counting\cite{ehucount,seatpeoplecounting}, people localization \cite{phasephi2,motion1,motion2,motion3}, vital sign detection\cite{respiration,repirationrate,respiratorypattern}, and gesture recognition \cite{falls1,falls2,rewis,activities,activity1,activitiy2,activity3,activity4,activity5}. Likewise, CSI-based sensing can also be employed in other applications, such as electrical device classification based on the effect of the impulsive noise in the received signals\cite{impulsivenoise}. In addition, it is worth noticing that IEEE 802.11 has recently approved a new task group named IEEE 802.11bf to accommodate sensing operations \cite{project} into the WiFi standards.

CSI measures the channel frequency response  (CFR) of a wireless communication link based on OFDM. This information is given in the form of amplitude and phase of the propagation channel for each  subcarrier in an OFDM symbol. While the CSI amplitude provides a reasonably accurate estimation of the CFR amplitude, the phase contains uncertainties that make it challenging to use in many applications and theoretical developments of HAR. Correcting these phase uncertainties in  frequency and time domains is a complex task, so many proposals in wireless sensing choose to work exclusively with amplitude \cite{ehucount,seatpeoplecounting,phasephi2,motion1,motion2,motion3,respiration,repirationrate,respiratorypattern,falls1,falls2,rewis,activities,activity1,activitiy2,activity3,activity4,activity5,impulsivenoise}. 

However, some authors have developed methods to perform channel phase estimation of WiFi CSI for indoor monitoring. A wide variety of these techniques perform a linear transformation (LT) of the phase to correct the linear impairments caused by synchronization issues. In this regard, LT is first used by Souvik Sen et al. \cite{monalisa}, where CSI is used to detect the position of people within different rooms, obtaining reasonable results. Qian, K. et al. \cite{pads} derive meaningful phase information by employing LT on the raw CSI to eliminate the significant random noise in the frequency domain. Outlier filtering is applied to shift out biased observations. Extracting various statistical features, such as variance, mean, and distribution distance, they obtain an accuracy of 90\% for human motion when using three antennas. Wang, X. et al. \cite{phasephi} use LT method for correcting the phase and then employ a deep network with three hidden layers to train the calibrated phases. Their results for detecting human positions in two rooms have about a 20\% of error in distance. Also, Fang, S. et al. \cite{coche} use LT to calibrate CSI phases, then an algorithm to extract different features is used, and last, a Deep Neural Network (DNN) classifies among three different human activities inside a car. In their work, Dang, X. et al. \cite{phase1} perform phase LT before using the difference between adjacent subcarriers to train a backpropagation neural network with fingerprint data. Recently, Cheng, X. et al.\cite{LT2021} constructed the phase difference matrix expanded by the mean and standard deviation of the phase difference as a feature matrix after the LT method. Then Savitzky-Golay filter is performed on the raw CSI phase information. More recently, Bu, Q. et al.\cite{LT2022} introduce TransferSense, a one-time, environment-independent WiFi sensing method based on deep learning that converts RF sensing tasks into image classification and uses amplitude and phase data corrected with the LT method. 

In addition to the LT method, other variants try to correct the errors of the estimated CSI phase, including non-linear errors. In the work of Kotaru, M. et al.\cite{spotfi}, a similar linear phase calibration method is developed as an extension of the LT processing to multiple antennas, using the frequency difference between subcarriers to estimate the phase. Zhu, H. et al. \cite{pisplicer} correct linear errors in phase, assuming that the CFR for one specific frequency should be the same even when measured in different bands. Then, they determine the time and frequency offsets in each band by matching the terms which define the CFR in each band. In Tadayon, N. et al. paper\cite{gauss}, the authors estimate time and frequency phase offsets separately. For time offsets, they assume that the channel impulse response is a linear combination of periodic functions whose period varies smoothly from sample to sample and try to correct the jumps observed at the power delay profile. For frequency offsets, the authors prove that the phase of the signal at the receiver follows a normal distribution to obtain an average value of the phase for all subcarriers in each packet or symbol. In a novel work, Meneghello, F. et al. \cite{new_dataset_tmc}, the authors consider a multipath propagation model for each CSI sample, using the most robust path as a reference to correct for time offsets in the remaining paths. They define each CSI sample as a product between the contributions that depend on the subcarriers index and a vector representing the independent terms from multipath. Then, for each subcarrier, they calculate the terms of that product and multiply it by the conjugate of the one with the strongest path to eliminate constant offsets in frequency.

After correcting the phase of the CSI, some works employ filtering to remove noise from the signal. One of the most recently used is the Savitzky-Golay (SG) filter, since it allows data smoothing with a reduced distortion of the signal tendency. This filter has been applied for wireless sensing in the frequency domain \cite{LT2021} and in the time domain \cite{wicount,sgtime}. 

In our previous work \cite{bmsb}, a channel phase calibration method was presented based on a linear regression of the CSI phase. In addition, time smoothing of the phase was carried out through a SG filter, and finally, an algorithm was proposed to correct phase gaps in frequency. The calibration method was tested using the power profile of simulated wireless channels. Based on the previous work, this manuscript presents an improved and extended method of phase processing for HAR in wireless sensing. Additionally, to validate the proposed method, a comprehensive analysis of the proposal is performed over five different datasets of experimental measurements of HAR, using three different neural networks and comparing it with five other types of phase processing.

Considering the above, this paper focuses on channel phase processing to improve the accuracy of HAR classification in indoor environments with OFDM-based wireless signals. The contributions of this manuscript to the current state of the art are the following:
\begin{itemize}
\item We propose a novel phase processing method of CSI, coined Time Smoothing and Frequency Rebuild (TSFR), to be used for HAR. It consists, first, of an improved model for channel phase sanitization \cite{bmsb}, adjusting and removing some parameters from the previous work. In addition, a new algorithm has been developed to smooth the phase in the time domain and correct discontinuities generated in frequency after filtering.
\item Two new CSI-based datasets with real measurements have been generated for counting people and position localization in indoor environments.
\item Two regular deep neural networks have been designed for CSI-based HAR: a fully connected network with four hidden layers and one dropout layer, and a convolutional network with three convolutional layers, three max pool layers and two flatten layers. In addition, the few-shot meta-learning technique named ProtoNet \cite{protonet} is also implemented to check the transferability of the results.
\item It presents a comprehensive performance analysis of the TSFR proposal for HAR purposes over five datasets (two new, three from the bibliography) and three deep learning models (two new, one from the bibliography). In this analysis, the use of the SG filter in the time domain, frequency domain, and in both domains simultaneously has been assessed. In addition, performance comparisons have been carried out with the other methods from the state of the art. Furthermore, performance results in terms of accuracy and confusion matrices have been obtained when working with the processed CSI phase, the CSI amplitude, and both variables combined.
\end{itemize}

The rest of the paper is organized as follows: Section \ref{sec:concepts} summarizes the main concepts on which our proposal is based: CSI and the Savitzky-Golay filter. Section \ref{sec:method} presents the proposed method. The datasets and the deep learning algorithms which are utilized in this work are described in Section \ref{sec:overview}. The results and discussion are presented in Section \ref{sec:results}. Finally, the paper will be concluded with some ideas and future directions in Section \ref{sec:conclusions}.

\textit{Notation}: Matrices are represented in capital letters and boldface. The matrix $\boldsymbol{O}_{SxK}$ represents a zero matrix with $S$ rows and $K$ columns. $\boldsymbol{O}_{*,k}$ represents the column vector $k$, and $\boldsymbol{O}_{s,*}$ the row column $s$. The application of the Savitzky-Golay filter is represented as $SG$\{$\cdot$ , $n$, 2$l$+1\} being $n$ the order of the polynomial used to fit the samples and $l$ the length of the filter window.
\section{Preliminary concepts}
\label{sec:concepts}
\subsection{Channel State Information}
CSI describes the properties of the channel through which the signal propagates, in this case, OFDM wireless signals. These channel properties depend on the environment and the propagation medium and can therefore be used to extract characteristics of the environment. In the field of HAR, CSI is widely used because the channel properties are affected by environmental changes. So these variations are associated with the different activities to be classified.

For an OFDM system, the received signal in the frequency domain can be modeled as
\begin{equation} \boldsymbol{y} = \boldsymbol{H}\cdot \boldsymbol{x}+ \boldsymbol{z} \label{y}\end{equation}

where $\boldsymbol{y}$ and $\boldsymbol{x}$ denote the received and transmitted signal vectors, respectively, $\boldsymbol{z}$ is the additive complex white Gaussian noise, and $\boldsymbol{H}$ represents a diagonal matrix of the CFR, also referred as CSI. 
The CSI of the $k$-th subcarrier during the $s$-th symbol, $h_{s,k}$, is a complex value as follows: 

\begin{equation}h_{s,k}=\left|h_{s,k}\right|e^{j\theta_{s,k}}\label{eq}\end{equation}
where $\left|h_{s,k}\right|$ and $\theta_{s,k}$ are the amplitude and the phase, respectively.
The CSI is therefore composed of two independent sources of information, amplitude on the one hand and phase on the other. 

At the receiver side, CSI is usually estimated to decode the received signal. In this process, synchronization issues can lead to several errors in the estimated CSI, making the treatment of the phase complex for HAR purposes due to its uncertainties and offsets. In particular, there are three main types of errors \cite{offsets2} affecting the phase that do not reduce communication quality but are of great importance when working with CSI for HAR classification in closed environments. 
\begin{itemize}
    \item Sample Frequency Offset (SFO) is due to a mismatch of the oscillators between the transmitter (TX) and the receiver (RX). This lack of synchronization generates a time shift of the received signal concerning the transmitted signal. As the local oscillator remains stable over a short time, the SFO is usually treated as a constant.
    \item Sample Time Offset (STO) occurs because the receiver detects the packet by correlation operation and signal power calculation. Due to hardware imperfection, this process introduces a random time shift.
    \item Carrier Frequency Offset (CFO) occurs because the receiver center frequency is not synchronized. The system completes the estimation and compensation at the receiver by analyzing the cyclic prefix and pilot signals. However, due to hardware instability, the frequency offset cannot be entirely determined, and this residual offset causes a non-negligible error in the phase.
\end{itemize}
Therefore, let $\boldsymbol{\widehat{H}}_{SxK}$ be the estimated CSI matrix of $S$ symbols with $K$ subcarriers:

\begin{equation}
\boldsymbol{\widehat{H}}_{SxK} =
\begin{pmatrix}
\widehat{h}_{1,1}& \cdots & \widehat{h}_{1,K}\\
\vdots & \ddots & \vdots\\
\widehat{h}_{S,1} & \cdots & \widehat{h}_{S,K}
\end{pmatrix}
\label{csi}
\end{equation}
where the $(s,k)$-th element of $\boldsymbol{\widehat{H}}$ can be given by $\widehat{h}_{s,k}= \left|\widehat{h}_{s,k}\right|e^{j\widehat{\theta}_{s,k}}$. Note that each row in Eq.\eqref{csi} corresponds to one diagonal in $H$ from Eq.\eqref{y}. Likewise, a matrix of the measured CSI phases can be defined as $\widehat{\boldsymbol{\Theta}}_{S \times K}$ where the measured phase at the $k$-th subcarrier of the $s$-th CSI frame can be expressed as:

\begin{equation}
\widehat{\theta}_{s,k}=\theta_{s,k}+\underbrace{2\pi\frac{m_k}{N}\cdot\Delta t}_{SFO,STO}+ \underbrace{\gamma}_{\\CFO}+Z\label{phase_measured}
\end{equation}
where $\theta_{s,k}$ is the actual phase, $\Delta t$ is the time lag due to SFO and STO, $m_k$ is the subcarrier index of the $k$th subcarrier, $N$ is the discrete Fourier transform size for the OFDM generation, $\gamma$ is the unknown phase offset due to CFO, and $Z$ is the measurement noise.
 
It is worth mentioning that these offsets occur in frequency and time domains and that SFO and STO linearly depend on each subcarrier. In the following, several phase processing methods are presented to provide useful information to the models used in the field of HAR. 

\subsection{Linear transformation}
\label{subsec:lt}
A usual approach to mitigate offset mismatches is to apply a linear transformation. 
It is noticed that the phase error $2\pi\frac{m_k}{N}\Delta t+\gamma$ in Eq.\eqref{phase_measured} is a linear function of the subcarrier index $m_k$. We can estimate for each symbol $s$ the phase slope $\varepsilon_s$ and the offset $\tau_s$ with the following expressions:

\begin{equation}\varepsilon_s=\frac{\widehat{\theta}_{s,K}-\widehat{\theta}_{s,1}}{m_K-m_1}\end{equation}
\begin{equation}\tau_s=\frac{1}{K}\sum_{k=1}^{K}{\widehat{{\theta}}}_{s,k}
\end{equation}

Finally, subtracting $\varepsilon_s m_k+\tau_s$ from the raw phase $\widehat{{\theta}}_{s,k}$, we can obtain the calibrated phase, ${\theta}^{\prime}_{s,k}$,  which is given by

\begin{equation}{\theta}^{\prime}_{s,k}=\widehat{{\theta}}_{s,k}-\ \varepsilon_s m_k-\tau_s\label{lt}\end{equation}

\subsection{Savitzky-Golay filter}
The Savitzky-Golay filter is a filtering method based on local area polynomial least square fitting for time-series signals \cite{Schafer2011}. It is used to smooth the CSI data and reduce environmental noise interference to facilitate the subsequent feature extraction \cite{SGWiFisensing}. The method requires defining a moving window of size $2l+1$ and a fitting order $n$ to perform left-to-right curve filtering. First, the filtering center is selected, and $2l+1$ point out of each $l$ point around the center is chosen as the primary filtering object. For the sake of simplicity, a vector $\boldsymbol{v}$ smoothed with a Savitzky-Golay filter is defined as
\begin{equation}
\label{SG}
\boldsymbol{v}_{sg}=SG\left(\boldsymbol{v},n,2l+1\right)
\end{equation}
where $\boldsymbol{v}_{sg}$ is the output of the filter.

The choice of the optimal parameters for the Savitzky-Golay filter depends on the nature of each problem. Therefore, they are obtained empirically through an analysis of data. In this work, the parameters of the one-dimensional SG filter have been obtained experimentally, being $n=2$ and $l$ is $0.1$ times the length of the input vector. These are the best values for obtaining the highest accuracy for data classification (counting and activities) with this paper’s evaluation methods.

Given a  CSI phase matrix $\widehat{\boldsymbol{\Theta}}$, SG filtering could be applied in every dimension of the matrix or in both at the same time. The CSI phase is expected to exhibit continuity across consecutive subcarriers of OFDM symbols because the channel coherence bandwidth has to be larger than the subcarrier spacing since the maximum delay spread must be much smaller than the symbol duration for the WiFi system to operate in a given environment. Furthermore, when CSI estimates of different symbols are obtained with a periodicity small enough compared with the coherence time of the wireless channel, continuity of the phase is also preserved in the inter-symbol time domain. 

Therefore, this work evaluates the SG filtering in order to maintain the continuous form of the phase in the frequency domain, in the inter-symbol time domain and in both domains as follows:
\begin{enumerate}
    \item Frequency domain: It is applied to the CSI estimate of each symbol. One of the filter characteristics is that it retains the width \& height of waveform peaks in noisy signal\cite{SG_stability}.
    \item Time domain: The filter is applied to each subcarrier along consecutive CSI symbols to smooth and ensure phase continuity over time. Its application has an impact on the frequency domain and can generate distortions.
    \item Time-Frequency domains: Applying the filter in both domains at the same time ensures continuity and phase smoothing in both dimensions. This data processing is done according to \cite{SG2d}.
\end{enumerate}

\section{Proposed method for phase processing}
\label{sec:method}
\subsection{Phase sanitization}
The idea of the proposed method is to take advantage of the good results offered by the linear transformation while maintaining continuity, i.e., avoiding gaps, in at least one of the two domains of the phase. In this sense, we improve the traditional linear transformation shown in Subsection \ref{subsec:lt} using a linear regression of overall symbol points to remove the slope generated by STO and SFO impairments. The amplitude, $\left|\widehat{h}_{s,k}\right|$, remains constant and unchanged throughout the entire phase sanitation process.

Given the phase matrix $\boldsymbol{\widehat{\Theta}}$, a linear regression of each $s$ symbol (i.e., $\widehat{\boldsymbol{\Theta}}_{s,*)}$ in $\boldsymbol{\widehat{\Theta}}$ is computed. Then, the linear regression model function follows the form:

\begin{equation}r_s(k)=\varepsilon_s\cdot k + \tau_s\label{rs}\end{equation}where $\tau_s$ is the offset:

\begin{equation}\tau_s=\boldsymbol{\bar{\widehat{\Theta}}}_{s,*}-\bar{k}\cdot \varepsilon_s\label{bs}\end{equation}and $\varepsilon_s$ is the linear regression slope:

\begin{equation}\varepsilon_s=\frac{\sum_{k=1}^{K}\left({\widehat{\theta}}_{s,k}-\bar{\widehat{\boldsymbol{\Theta}}}_{s,*}\right)\left(k-\bar{k}\right)}{\sum_{k=1}^{K}\left(k-\bar{k}\right)^2}\label{as}\end{equation}
Note that $\boldsymbol{\bar{\widehat{\Theta}}_{s,*}}$ and $\bar{k}$ are the average value of $\boldsymbol{\widehat{\Theta}}_{s,*}$ and $k$, respectively.

Finally, following the shape of Eq.\eqref{lt}, we can obtain the corrected phased $\widecheck{\theta}_{s,k}$ as:
\begin{equation}
\widecheck{\theta}_{s,k}=\widehat{\theta}_{s,k} - k \cdot \varepsilon_s - \tau_s
\label{rot}
\end{equation}

Figure \ref{lrr_tsfr}a shows a graphical example of the LT method and the proposed Linear Regression Transformation (LRT) solution for a specific CSI frame in the OPERAnet dataset \cite{operanet}.

\subsection{Time Smoothing and Frequency Rebuild}
\label{subsec:TSFR}

\begin{figure}[b]
\centering
\begin{subfigure}[b]{0.23\textwidth}
    \centering
    \includegraphics[width=\textwidth]{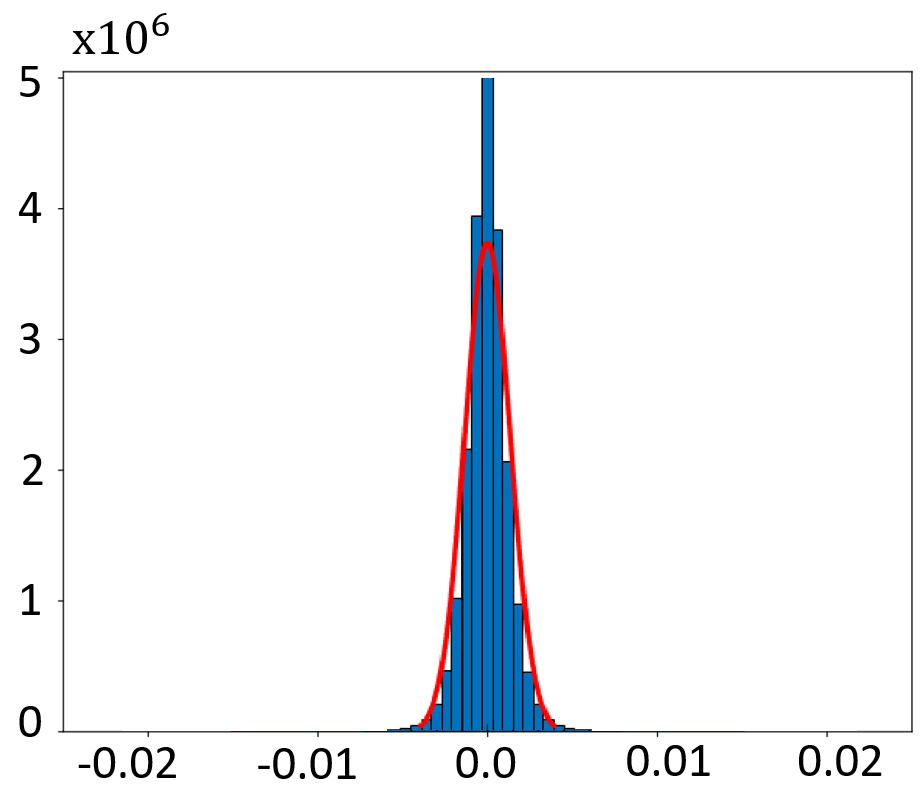}
    \caption{Dataset A}
\end{subfigure}
\vspace*{\fill}
\begin{subfigure}[b]{0.24\textwidth}
    \centering
    \includegraphics[width=\textwidth]{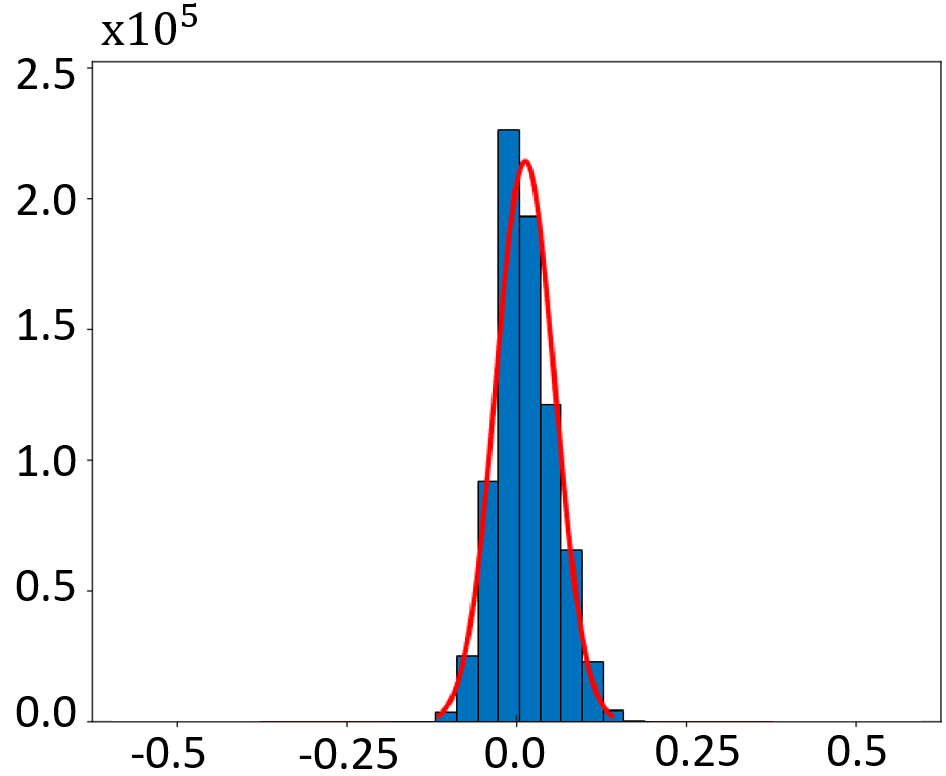}
    \caption{Dataset B}
\end{subfigure}
\vspace*{\fill}
\begin{subfigure}[b]{0.24\textwidth}
    \centering
    \includegraphics[width=\textwidth]{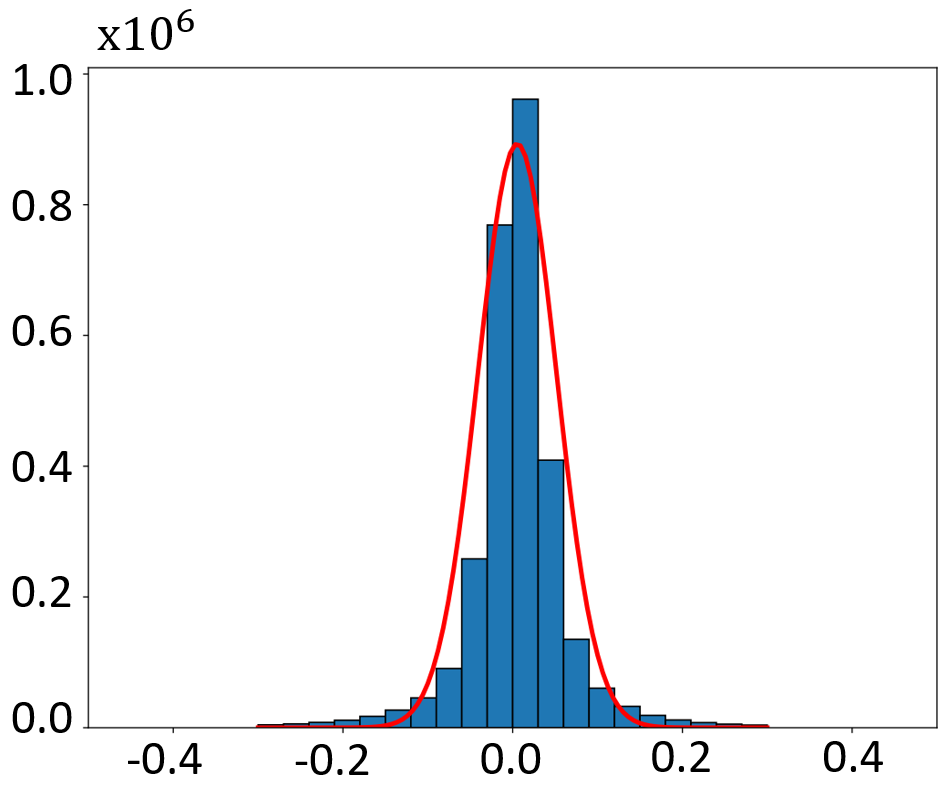}
    \caption{EHUCount}
\end{subfigure}
\begin{subfigure}[b]{0.23\textwidth}
    \centering
    \includegraphics[width=\textwidth]{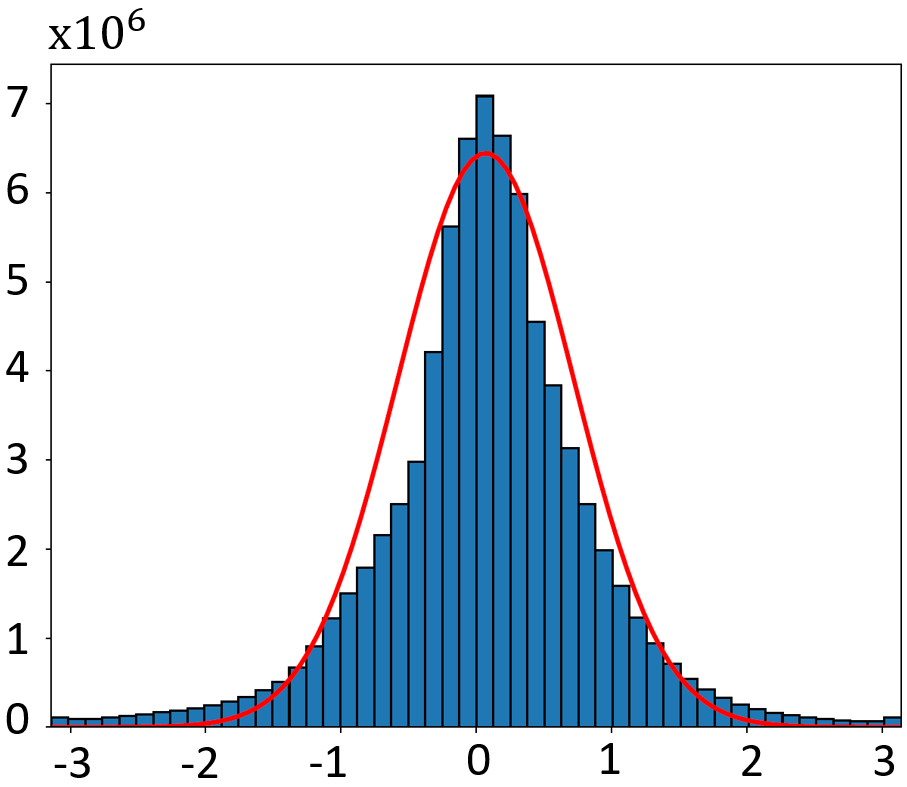}
    \caption{OPERAnet}
\end{subfigure}
\caption{Histograms and Gaussian distribution approximations of $\boldsymbol{\widecheck{\theta}}_{s,k}-{\boldsymbol{\widecheck{\theta}}}_{s,k-1}$ for HAR datasets described in Section \ref{sec:overview}.} 
\label{diff_ang}
\end{figure}

At this point, calibrated phases maintain distortions and gaps between adjacent CSI symbols in $\widecheck{\boldsymbol{\Theta}}$. Moreover, applying LT or LRT methods cannot ensure phase continuity in frequency, since other non-linear errors in hardware, software, or a weak implementation of the measurements can also generate gaps and deform the received signal and, in consequence, the estimated CSI. For this reason, a low-pass filter is used to smooth the calibrated CSI phases and ensure phase continuity. Time domain gaps correction makes sense if the activity to be measured generates changes in the channel at a rate greater than the time interval between OFDM symbols, which is the case in this work and generally in the HAR field.

Time Smoothing and Frequency Rebuild, TSFR, is the method proposed in this section. Assuming that the calibrated phase is approximately continuous in frequency and the main discontinuities appear in the time domain between adjacent symbols, SG filtering is proposed to be applied in the time domain, combined with a threshold-based method to correct the irregularities that SG filtering generates in the frequency domain and, thus, to maintain continuity.

Once $\boldsymbol{\widecheck{\Theta}}_{SxK}$ have been calculated, SG filtering is carried out in the time domain subcarrier as:

\begin{equation}
\boldsymbol{\widecheck\Phi}_{*,k}=SG\left(\boldsymbol{\widecheck{\Theta}}_{*,k},\ 2,\ 0.1\cdot S\right)
\label{SG2}
\end{equation}

Due to the previous time filtering, discontinuities in the frequency domain are generated in the form of a step between subcarrier blocks. A threshold-based method is proposed to remove those quantitatively large gaps that can appear between two adjacent subcarriers. In most scenarios, it can be assumed that the phase of the channel frequency response change slowly between adjacent subcarriers. As a result, the phase difference between adjacent subcarriers (i.e., $\widecheck{\theta}_{s,k}-{\widecheck{\theta}}_{s,k-1}$) should be small and large gaps could be considered outliers. Considering that those noisy differences can be approximated to a Gaussian distribution \cite{gauss}, we have defined a threshold, $d_s$, which has the form:
\begin{equation}
d_s=\mu_s+\sigma_s
\label{ds}
\end{equation}
where $\mu_s$ is the average of the phase differences before SG filtering:
\begin{equation}
\mu_s=\frac{1}{K-1}\sum_{k=2}^{K}\left|\widecheck{\theta}_{s,k}-{\widecheck{\theta}}_{s,k-1}\right|\ \ \ 
\label{mu}
\end{equation}
and $\sigma_s$ is the standard deviation:
\begin{equation}
\sigma_s=\ \sqrt{\frac{\sum_{k=2}^{K}\left(\left|\widecheck{\theta}_{s,k}-{\widecheck{\theta}}_{s,k-1}\right|-{\ \mu}_s\right)^2}{K-1}}.
\label{sigma}
\end{equation}

\begin{figure}[t]
\centering
\begin{subfigure}[b]{0.25\textwidth}
    \centering
    \includegraphics[width=\textwidth]{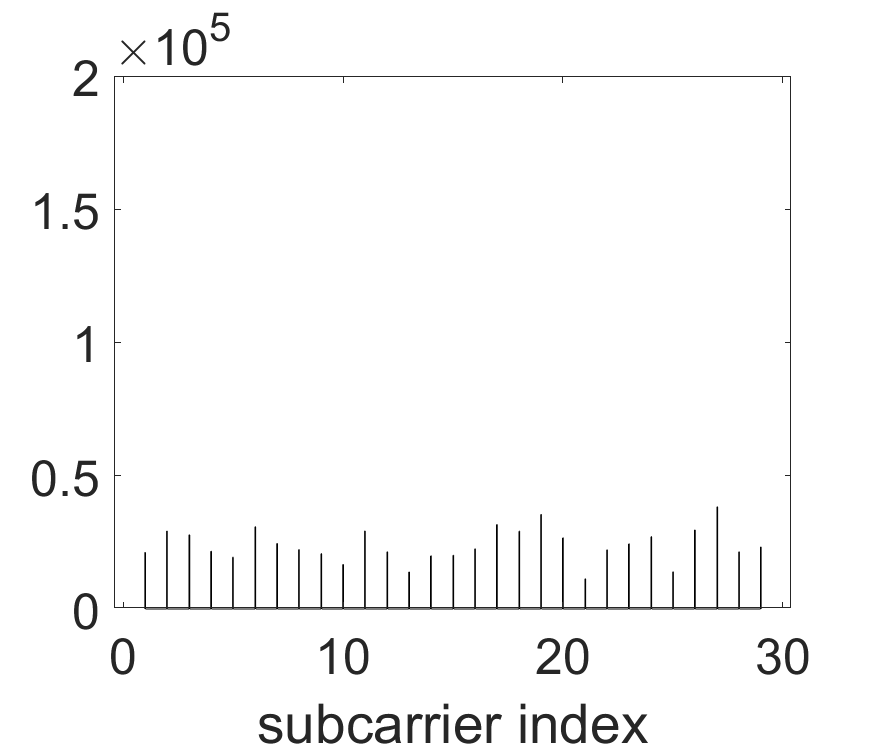}
    \caption{Empty}
\end{subfigure}
\begin{subfigure}[b]{0.25\textwidth}
    \centering
    \includegraphics[width=\textwidth]{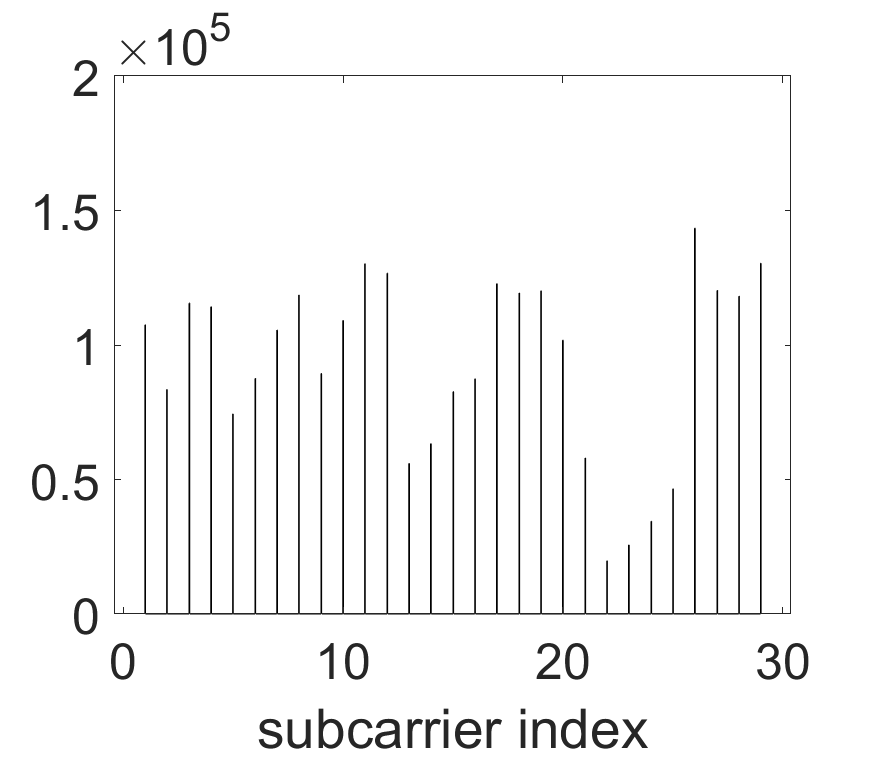}
    \caption{1 person}
\end{subfigure}
\begin{subfigure}[b]{0.25\textwidth}
    \centering
    \includegraphics[width=\textwidth]{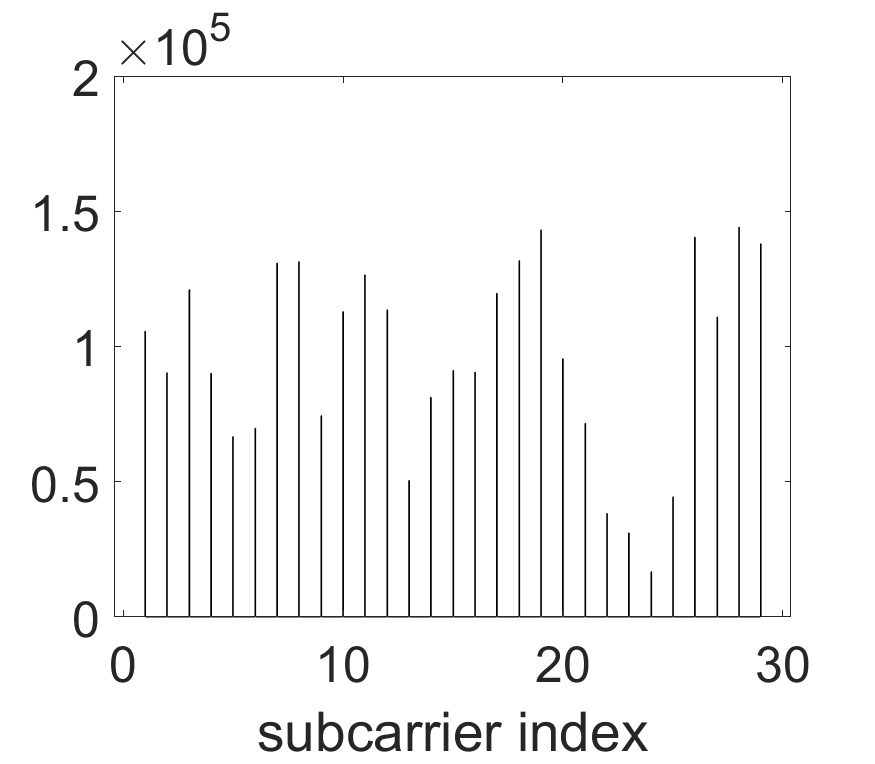}
    \caption{2 people}
\end{subfigure}
\begin{subfigure}[b]{0.25\textwidth}
    \centering
    \includegraphics[width=\textwidth]{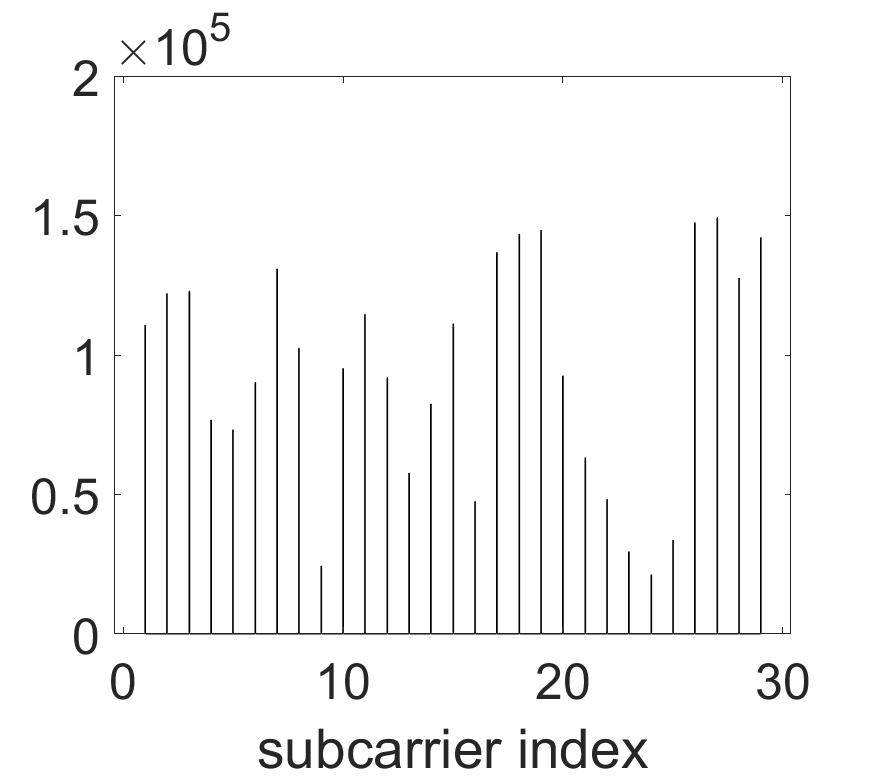}
    \caption{3 people}
\end{subfigure}
\begin{subfigure}[b]{0.25\textwidth}
    \centering
    \includegraphics[width=\textwidth]{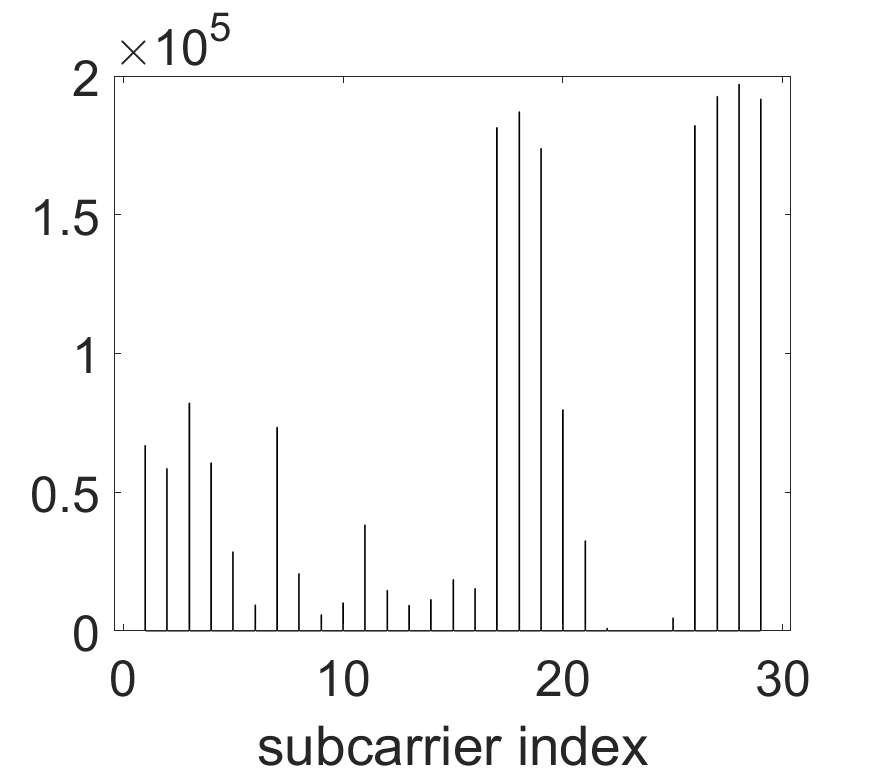}
    \caption{4 people}
\end{subfigure}
\begin{subfigure}[b]{0.25\textwidth}
    \centering
    \includegraphics[width=\textwidth]{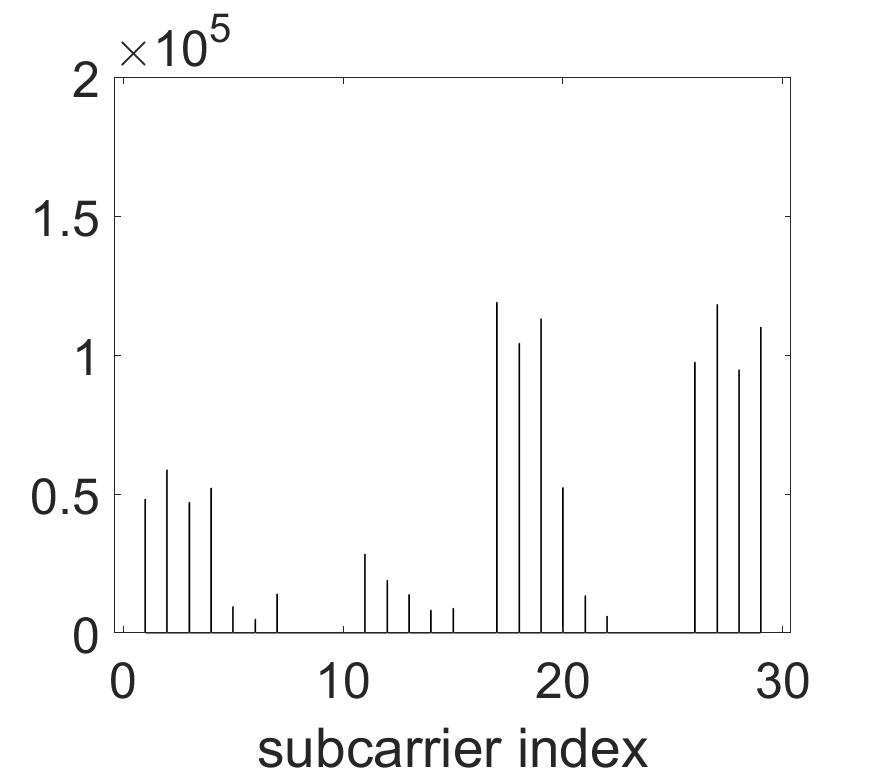}
    \caption{5 people}
\end{subfigure}
\caption{Frequency distributions of the subcarriers, $k$, on which condition $|\widecheck{\phi}_{s,k}-{\widecheck{\phi}}_{s,k-1}| > d_s$ is satisfied for different number of people in a given room. These data belong to the OPERAnet dataset: tx3rx3.}
\label{fig:k_index}
\end{figure}

Consequently, after SG filtering in the time domain, a TSFR phase matrix $\boldsymbol{\widetilde{\Phi}}_{SxK}$ is calculated where the $(s,k)$-th element follows:

\begin{equation}
\widetilde{\phi}_{s,k}=\left\{\begin{array}{ll}
    \widecheck{\phi}_{s,k}\ &  \text{if\ } k=1\\
    \widetilde{\phi}_{s,k-1} - d_s \ &  \text{if\ } \ \epsilon<-d_s  \\
    \widetilde{\phi}_{s,k-1} + d_s \ &  \text{if\ } \ \epsilon>d_s  \\
    \widecheck{\phi}_{s,k}-(\widecheck{\phi}_{s,k-1}-\widetilde{\phi}_{s,k-1})  & \ \text{otherwise }\\ 
\end{array} \right.
\label{phifinal}
\end{equation}being $\epsilon = \widecheck{\phi}_{s,k}-\widecheck{\phi}_{s,k-1}$.

In Fig. \ref{diff_ang}, several histograms of the phase differences of adjacent subcarriers are drawn for different datasets, which are described in Section \ref{sec:overview}. One can observe that phase difference distributions present a bell-shape and can be approximated to a Gaussian distribution.

\begin{figure*}[t]
\centering
\begin{subfigure}[b]{0.19\textwidth}
    \centering
    \includegraphics[width=\textwidth]{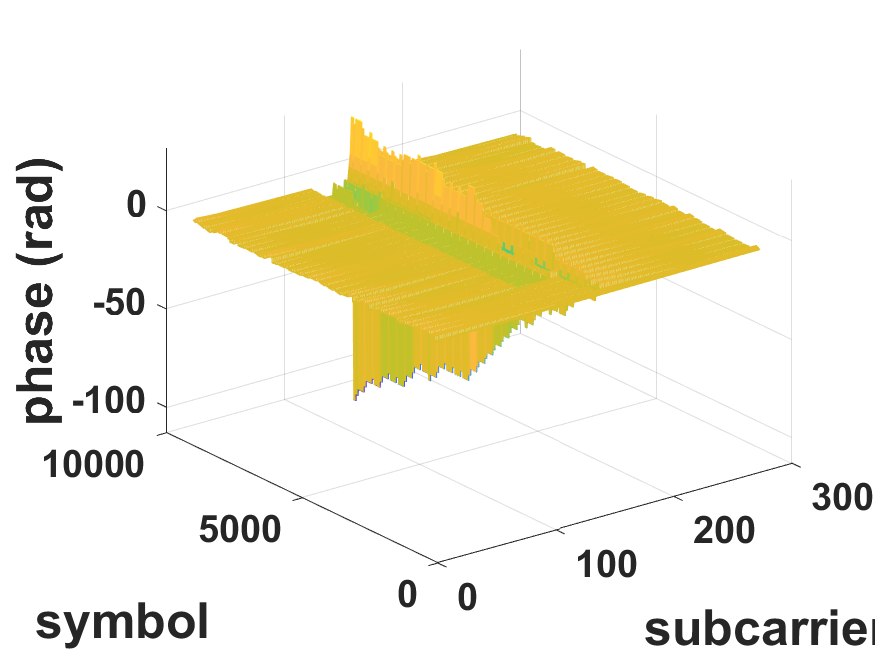}
    \caption{Raw}
\end{subfigure}
\begin{subfigure}[b]{0.19\textwidth}
    \centering
    \includegraphics[width=\textwidth]{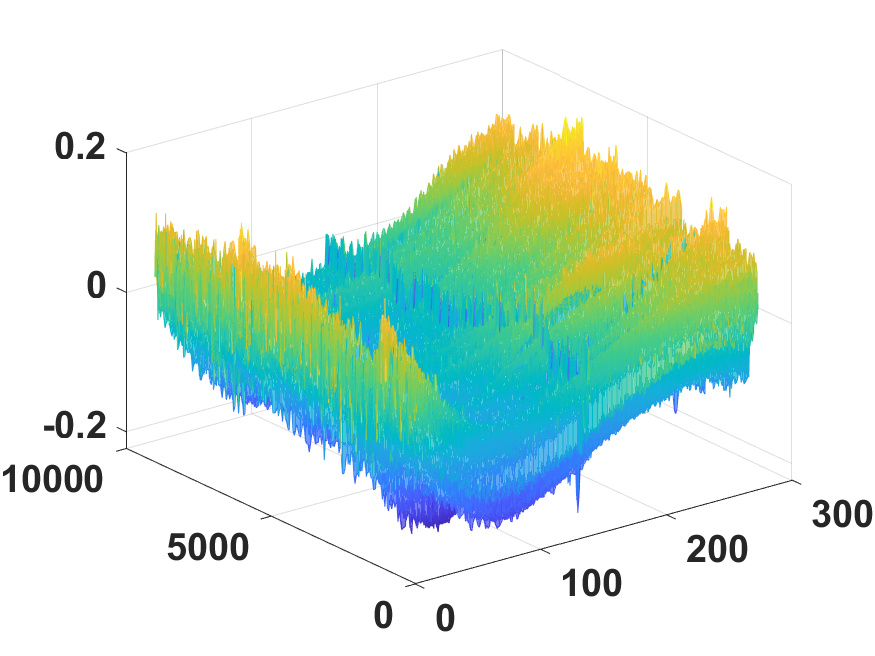}
    \caption{LT}
\end{subfigure}
\begin{subfigure}[b]{0.19\textwidth}
    \centering
    \includegraphics[width=\textwidth]{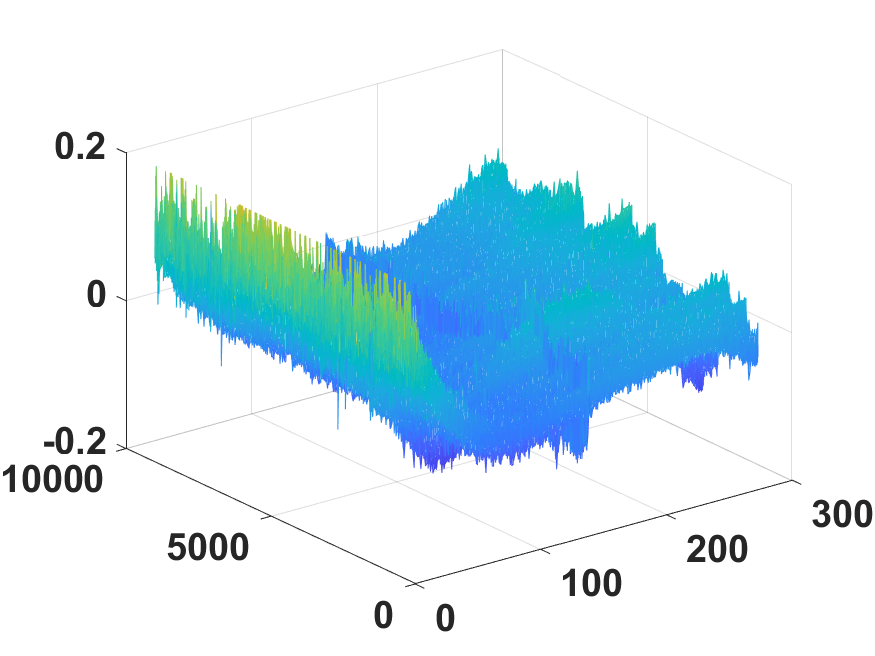}
    \caption{LRT}
\end{subfigure}
\begin{subfigure}[b]{0.19\textwidth}
    \centering
    \includegraphics[width=\textwidth]{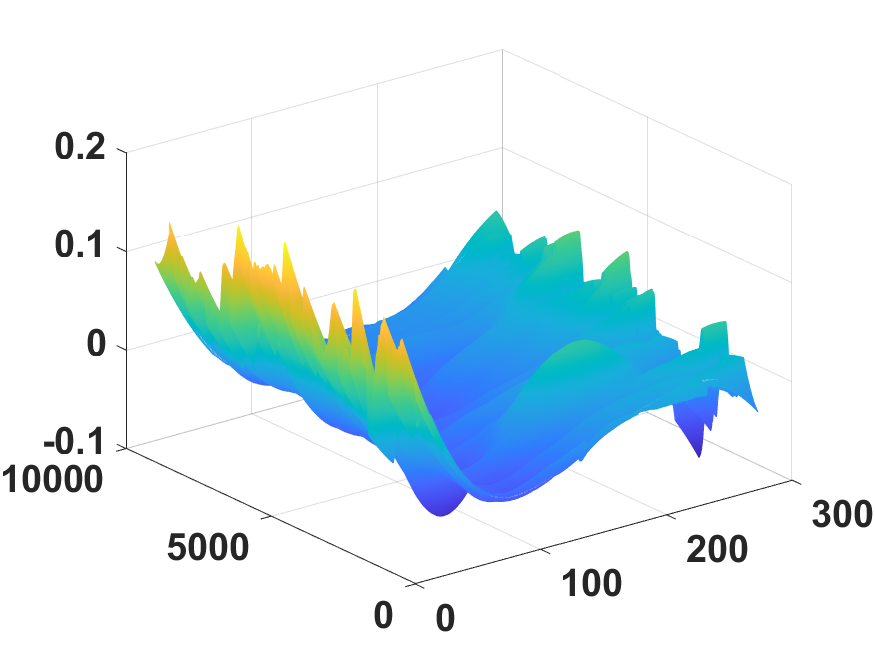}
    \caption{LRT+2D}
\end{subfigure}
\begin{subfigure}[b]{0.19\textwidth}
    \centering
    \includegraphics[width=\textwidth]{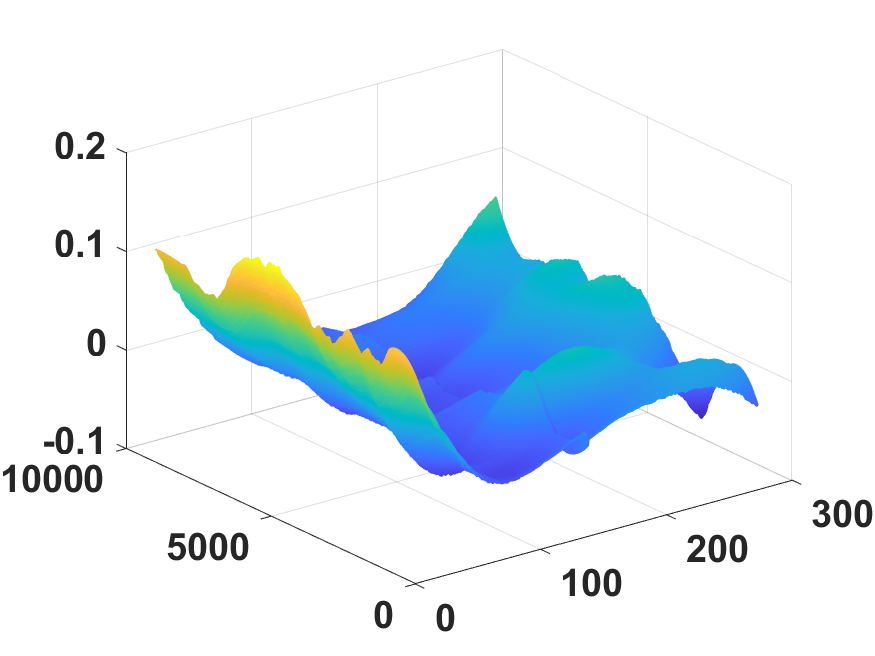}
    \caption{TSFR}
\end{subfigure}
\hss
\caption{Processed phase of CSI matrices with LT, LRT, LRT+2D SG filtering, and TSFR, using Dataset A.}
\label{dvbimage}
\end{figure*}

According to this methodology, the phase of any subcarrier in which the difference with the previous one exceeds $d_s$ will be modified. Based on the Gaussian assumption, approximately 30\% of the subcarriers of each symbol are modified, including outliers generated by the SG filtering and actual smoothed values. Therefore, this methodology is not only intended to correct the outliers due to temporal filtering. It also tries to take advantage of this correction to modify the statistical distribution of the symbol, making it more characteristic for each activity by means of the $d_s$ value. The main ideas behind this method are:
\begin{itemize}
    \item[--] The time evolution of the phase for each subcarrier can reveal information related to the channel variations of each activity. Therefore, those subcarriers that suffer phase gaps after time filtering are also characteristics of the time evolution of the whole CSI phase matrix, as shown in Fig. \ref{fig:k_index}. One can observe that some phase differences can be sensitive to the channel changes related to the activity in the room, while others behave steady.
    \item[--]The corrected phases after the LRT method in each CSI symbol also contain relevant information related to the channel variations of each activity. Part of this information is present in its statistical variables, such as those referred to Eq.\eqref{ds}, Eq.\eqref{mu} and Eq.\eqref{sigma}. The gaps generated as a result of the temporal filtering can corrupt this valuable information for HAR and, therefore, these gaps are reduced through the proposed adjustment in Eq.\eqref{phifinal}.
\end{itemize}

\begin{algorithm}
\begin{multicols}{2}
\caption{TSFR}
\label{alg1}
\textbf{Input:} $\boldsymbol{\widehat{\Theta}}_{SxK}$;\\
\vspace{0.25em}
\textbf{Output:} $\boldsymbol{\widetilde{\Phi}}_{SxK}$;\\
\vspace{1em}
$\boldsymbol{\widecheck{\Theta}}_{SxK}$ = $\boldsymbol{O}_{SxK}$; \\
\vspace{0.25em}
$\boldsymbol{\widecheck{\Phi}}_{SxK} = \boldsymbol{O}_{SxK}$; \\
\vspace{0.25em}
$\boldsymbol{\widetilde{\Phi}}_{SxK} = \boldsymbol{O}_{SxK}$; \\
\vspace{0.25em}
\vspace{1em}
\For{$s$ in 1:$S$}{
\vspace{0.25em}
    $\boldsymbol{\widehat{\theta}}_{s}$ $\leftarrow$ unwrap phase of $\boldsymbol{\widehat{\Theta}}_{s,*}$;\\
    \vspace{0.25em}
    $b_s$ $\leftarrow$ apply Eq.\eqref{bs};\\
    \vspace{0.25em}
    $a_s$ $\leftarrow$ apply Eq.\eqref{as};\\
    \vspace{0.25em}
    $\boldsymbol{\widecheck{\theta}}_{s}$ $\leftarrow$ apply Eq.\eqref{rot};\\
    \vspace{0.25em}
    $\boldsymbol{\widecheck{\Theta}}_{s,*} = \boldsymbol{\widecheck{\theta}}_{s}$;\\
}

\For{$k$ \text{in} 1:$K$}{
\vspace{0.25em}
    $\widecheck{\boldsymbol{\theta}}_{k}$ $\leftarrow$ unwrap phase of $\widecheck{\boldsymbol{\Theta}}_{*,k}$;\\
    \vspace{0.25em}
    $\widecheck{\boldsymbol{\phi}}_{k}$ $\leftarrow$ smooth $\widecheck{\boldsymbol{\theta}}_{k}$ applying Eq.\eqref{SG2};\\
    \vspace{0.25em}
    $\widecheck{\boldsymbol{\Phi}}_{*,k}$ = $\boldsymbol{\widecheck\phi}_{k}$
}
\For{$s$ \text{in} 1:$S$}{
\vspace{0.25em}
    $\widecheck{\boldsymbol{\phi}}_{s}$ $\leftarrow$ unwrap $\widecheck{\boldsymbol{\Phi}}_{s,*}$;\\
    \vspace{0.25em}
    $\widecheck{\boldsymbol{\theta}}_{s}$ $\leftarrow$ unwrap phase of $\widecheck{\boldsymbol{\Theta}}_{s,*}$\\
    \vspace{0.25em}
    $\mu_s$ $\leftarrow$ apply Eq.\eqref{mu};\\
    \vspace{0.25em}
    $\sigma_s$ $\leftarrow$ apply Eq.\eqref{sigma};\\
    \vspace{0.25em}
    $d_s$ $\leftarrow$ apply Eq.\eqref{ds};\\
    \vspace{0.25em}
    $\widetilde{\boldsymbol{\phi}}_{s}$ $\leftarrow$ apply Eq.\eqref{phifinal};\\
    \vspace{0.25em}
    $\widetilde{\boldsymbol{\Phi}}_{s,*} = \boldsymbol{\widetilde{\phi}}_{s}$;\\
}
\end{multicols}
\end{algorithm}
With this in mind, the $d_s$ value incorporates information related to each activity into the CSI phase matrix generating a characteristic modal number ($d_s$) for each symbol and applying it in the time domain via time characteristic subcarriers $k$, on which condition $|\widecheck{\phi}_{s,k}-{\widecheck{\phi}}_{s,k-1}| > d_s$ is satisfied. With this phase correction method, the jumps are not completely eliminated, but their value is reduced and uniformed to the $d_s$ value. So the information is preserved while distortion is reduced. In short, new information is added to each OFDM symbol using the new variable $d_s$: how many times it is repeated, between which subcarriers, and what its magnitude is. All this is intended to help the prediction algorithms to classify correctly. The benefits of this processing are confirmed by the good results obtained, as seen in section \ref{sec:results}.

\begin{figure*}[t]
\centering
    \begin{subfigure}[b]{0.5\textwidth}
        \centering
        \includegraphics[width=\textwidth]{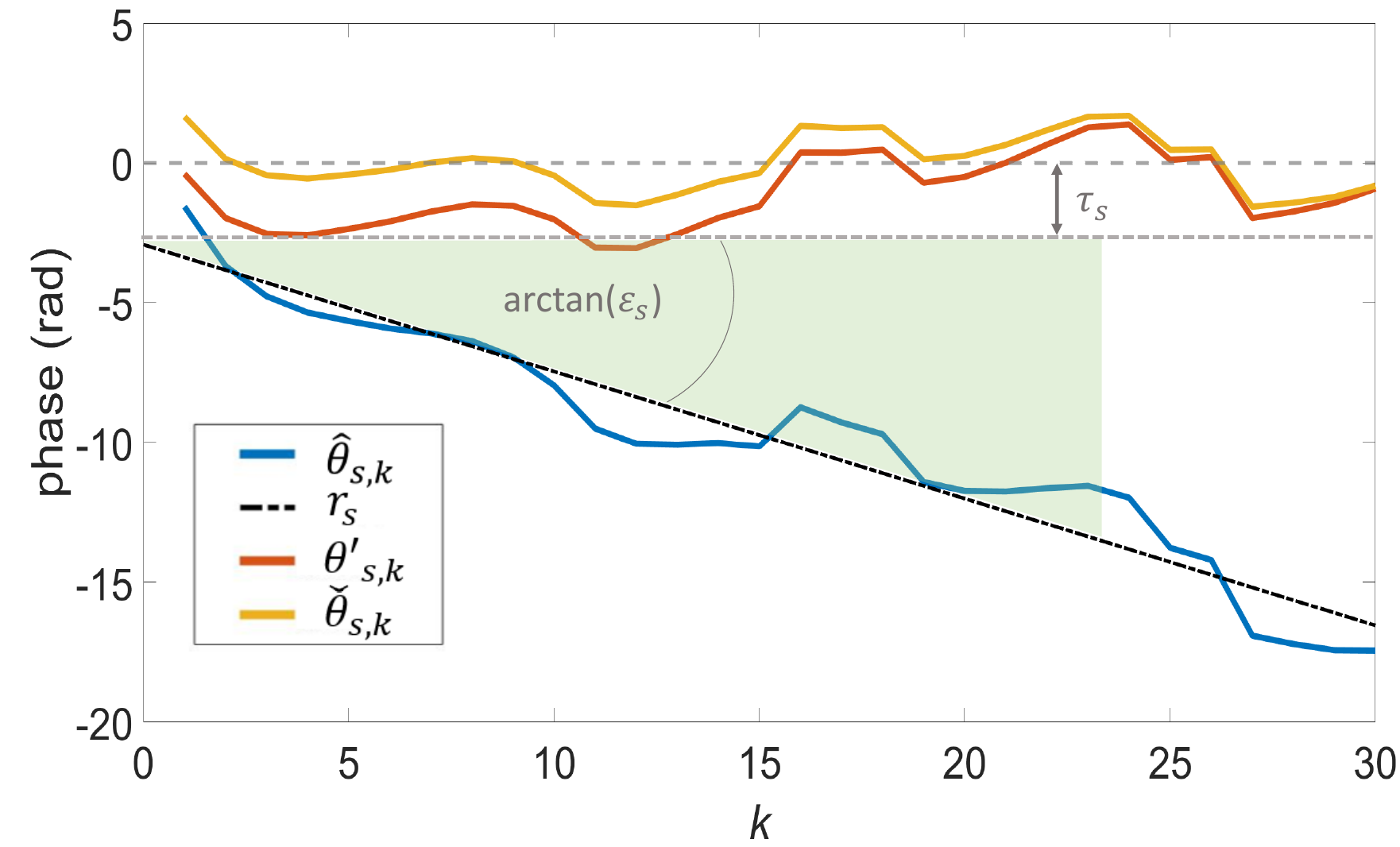}
        \caption{LRT}
    \end{subfigure}
    \hfill
    \begin{subfigure}[b]{0.495\textwidth}
        \centering
        \includegraphics[width=\textwidth]{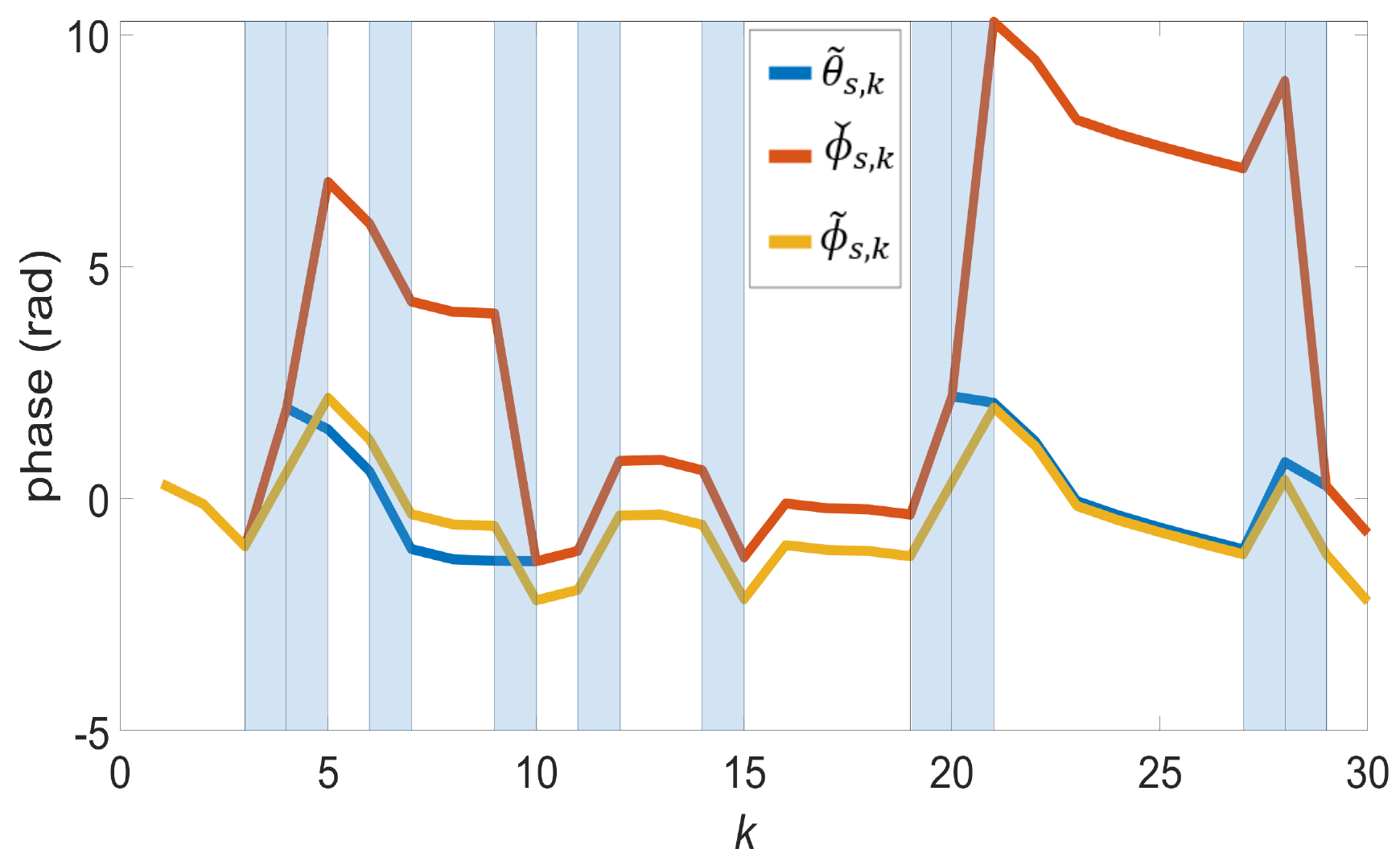}
        \caption{TSFR}
    \end{subfigure}
\caption{\textbf{(a)} Graphical representation of a CFO, SFO, and STO correction in a symbol phase by LRT ${\widecheck{\theta}}_{s,k}$ vs. bibliographic method ${\theta}^{\prime}_{s,k}$. These data belong to the OPERAnet dataset: tx1rx1, $s$ = 500K. \textbf{(b)} Phases after LRT (blue), time SG smoothing (red), and TSFR (yellow). Blue areas show adjacent subcarriers where $|\widecheck{\phi}_{s,k}-{\widecheck{\phi}}_{s,k-1}| > d_s$ is satisfied. These data belong to the OPERAnet dataset: tx1rx1, $s$ = 50. }
\label{lrr_tsfr}
\end{figure*}

In Fig. \ref{lrr_tsfr}b, the effect of the time SG filtering and the gap removal process is shown for a certain CSI symbol in the OPERAnet dataset (tx1rx1, $s$ = 50). We can observe that several large steps are generated after SG filtering in the blue areas and, afterward, removed with the proposed threshold-based method.

The complete TSFR method is described in Algorithm \ref{alg1}. In Fig. \ref{dvbimage}, representations of the processed CSI phase matrices using different phase processing are shown for a certain estimated phase CSI matrix $\boldsymbol{\widehat{\Theta}}_{SxK}$ corresponding to real measurements. One can initially observe the synchronization errors in the measured phases. Corrections of the linear phase impairments are carried out with the proposed LRT solution, and $\boldsymbol{\widecheck{\Theta}}_{SxK}$ matrix is depicted in Fig. \ref{dvbimage}c. In Fig. \ref{dvbimage}b, we can also observe the corrections performed with the traditional LT method. Finally, the output of the TSFR solution $\boldsymbol{\widetilde{\Phi}}_{SxK}$ is given in Fig. \ref{dvbimage}e. Additionally, we can see in Fig. \ref{dvbimage}d the processed phase when the LRT method is applied along with two-dimensional SG filtering.

Finally, after the TSFR-based phase processing, the processed CSI matrix $\boldsymbol{\widetilde{H}}_{SxK}$ can be reconstructed:

\begin{equation}
\boldsymbol{\widetilde{H}}_{SxK} =
\begin{pmatrix}
\widetilde{h}_{1,1}& \cdots & \widetilde{h}_{1,K}\\
\vdots & \ddots & \vdots\\
\widetilde{h}_{S,1} & \cdots & \widetilde{h}_{S,K}
\end{pmatrix}
\end{equation}where $\widetilde{h}_{s,k}=\left|\widehat{h}_{s,k}\right|e^{j\widetilde{\phi}_{s,k}}$.

\section{Evaluation Setup Description}
\label{sec:overview}

\subsection{Datasets}
This section explains the datasets used to test the proposed phase processing method for different human activity recognition (i.e., people counting people, position detection) and in indoor environments. There are five datasets. Two of them (named A and B) are not published and are available under request. The other three are public, and their characteristics are described in detail in their respective papers\cite{ehucount, operanet, rewis}.  The main characteristics of the datasets are shown in Table \ref{datasetstable} and  are explained below, especially for the ones that are not publicly accessible:
\begin{table}[h]
\caption{Datasets}
\centering
\label{datasetstable}
\setlength{\tabcolsep}{3pt}
\begin{tabular}{|p{50pt}|p{60pt}|p{110pt}|p{40pt}|p{25pt}|} 
\hline 
Name&	
System&	
Class&	
Scenarios&	
RXs\\
\hline
Dataset A&
DVB-T2 based&
Counting and fixed position&
1&
2\cr
Dataset B&
WiFi&
Counting and fixed position&
1&
1\cr
EHUCount&
WiFi&
Counting&
5&
1\cr
OPERAnet&
WiFi&
Counting&
1&	
3\cr
ReWiS&
WiFi&
Activities&
3&
1\cr
\hline 
\end{tabular}
\label{tab1}
\end{table}

\subsubsection{Dataset A}
Dataset A is a dataset created by our research group at the University of the Basque Country. The dataset's purpose was to count people and detect their fixed positions (sitting) in an indoor environment. The measurements were taken in a meeting room (2.8m x 4.8m) with one TX and two RXs. There were four chairs around a table in the room. The TX and the RXs were at the same height as the table. Measurements were taken in the presence of zero to four people. For each number of people, all the possible occupancy of the chairs were measured, e.g., with two people in the room, six measurements were performed, each time occupying different chairs. This makes a total of 16 measurements. The set-up for the measurements is sketched in Fig. \ref{scen}A.

Three USRPs (Universal Software Radio Peripheral) were used for these measurements, one as a TX and two as RXs. A DVB-T2, 10 MHz, 32K Digital Terrestrial Television (DTT) based signal was employed for the measurements. The channel frequency was 5.4 GHz, and the sampling frequency of the TX and the RX USRPs was doubled to obtain 20 MHz bandwidth (BW). A software (SW) DVB-T2 receiver was used to decode the T2 signal and obtain the CSI, which were then decimated to work with $K$=273 subcarriers at a rate of 606 Hz.

\subsubsection{Dataset B}
\begin{figure*}[hbt]
\centering
    \begin{subfigure}[b]{0.48\textwidth}
        \centering
        \includegraphics[width=\textwidth]{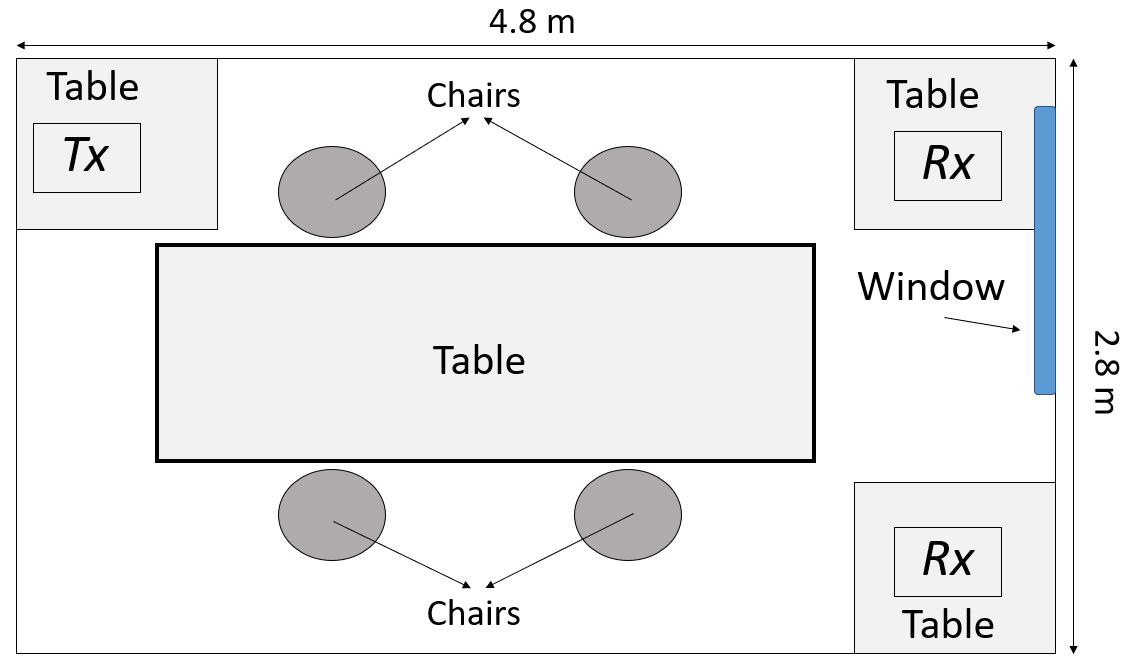}
        \caption{Dataset A}
    \end{subfigure}
    \hfill
    \begin{subfigure}[b]{0.48\textwidth}
        \centering
        \includegraphics[width=\textwidth]{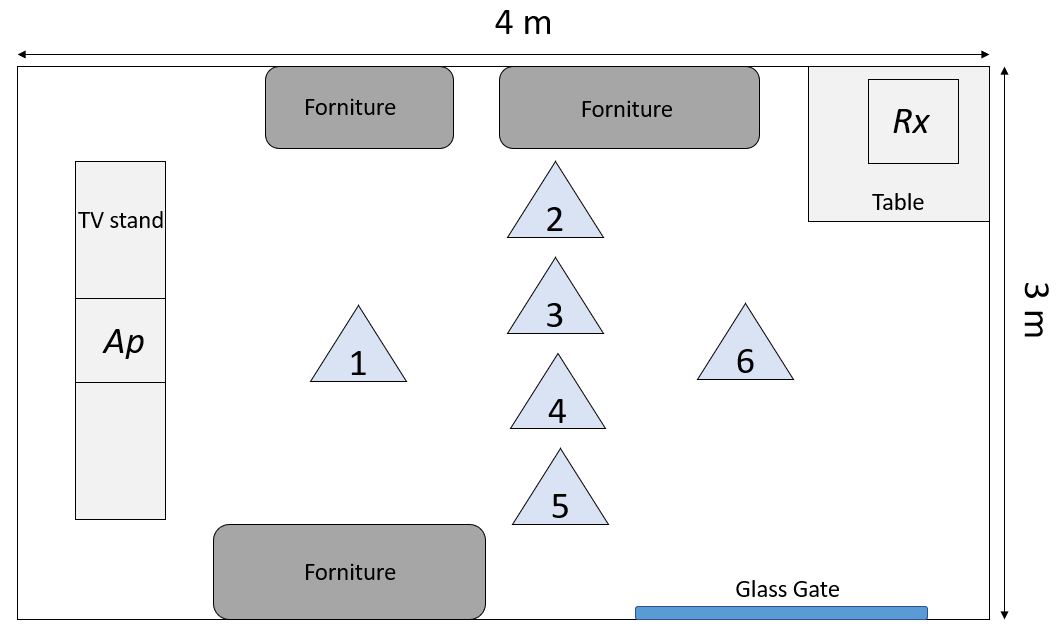}
        \caption{Dataset B}
    \end{subfigure}
\caption{Experimental rooms: \textbf{(a)} Dataset A and \textbf{(b)} Dataset B}
\label{scen}
\end{figure*}

This dataset was created by researchers at the National Autonomous University of Mexico (UNAM). The measurements were taken in the living room of the researcher's  apartment (approximately 3x4 meters). Six different locations  were selected in the room, and a chair was placed in each location.  The measurements were made with one or two people in the room, sitting on the chairs or standing in from of them, covering all the possible combinations of locations, number of people, and in sitting or standing positions. Measurements of the room without people were also taken, but only 1\% of the measurements correspond to this situation in contrast with 50\% of measurements with one person or 49\%  of two people, so the dataset is  strongly unbalanced. The room and the locations of the chairs are shown in Fig. \ref{scen}B.
 
The measurement system consisted of two laptops with Qualcomm Atheros QCWB335 network interface cards (NIC). One of the laptops injected WiFi packets, and the other received the signal and recorded the CSI. The Atheros-CSI-Tool \cite{Atheros_CSI} was used for this purpose.
Channel 11 of the 2.4 GHz WiFi band was used with 20 MHz BW (56 subcarriers). An average number of 50K packages were recorded, and the measurement time ranged from 13 to 18 seconds.

\subsubsection{EHUCount}
This dataset was obtained from measurements taken at the facilities of the Faculty of Engineering of the University of the Basque Country (Bilbao, Spain). The portable test bench consisted of a vector signal generator that was used to transmit a 15 s long pre-recorded IEEE 802.11n trace in the 2.4 GHz band with 20 MHz BW. The reception was performed by recording the signal as IQ samples with a signal analyzer to obtain the CSI using an SW WiFi demodulator. This dataset provides  CSI from $K$ = 52 subcarriers of the OFDM signal. 

Measurements were carried out in five indoor scenarios where up to five people walked casually. The number of CSI traces per number of people and scenario ranged between 12K and 15K, depending on synchronization issues in the signal decoding process.

\subsubsection{OPERAnet}
It is a comprehensive dataset intended to evaluate passive HAR and localization techniques with measurements obtained from synchronized Radio-Frequency devices and vision-based sensors. For our purposes, the dataset consists of CSI data extracted from a WiFi NIC.
Of the vast number of measurements and experiments in this dataset, we  only used one, named "exp028: Crowd counting". The "exp028" dataset contains the CSI from the three TX antennas to each of the three RX antennas. For example, the CSI matrix generated between TX antenna two and RX antenna two is called tx2rx2. For convenience, only tx1rx1, tx2rx2, and tx3rx3 data have been used in this work. 

For the experiment, a maximum of six people walked continuously and randomly through a room. It started with six people; then, every 5 minutes, one person left the monitoring area. 
The WiFi CSI system consisted of three PCs fitted with an Intel5300 NIC, which extracts CSI from $K$= 30 subcarriers, spread evenly among the 56 subcarriers of the 20 MHz channel 149 in the 5 GHz band at a rate of 1.6 kHz.

\subsubsection{ReWiS}
These measurements were carried out in three different settings. The experiments involved two subjects who were given instructions on the type, duration, and location of activities such as jumping, walking, and standing. Each measurement campaign involved 180 seconds of data collection for each activity performed by the two people. Measurements were repeated ten times with a time interval of at least 2 hours between measurements.
For the generation of the ReWiS dataset, the authors used three Asus RT-AC86U WiFi routers, each equipped with four antennas. The routers extracted the CSI packets using the Nexmon firmware \cite{nexmon}. The CSIs were calculated at a rate of 100 Hz, in the 5 GHz band, for 20 and 80 MHz BW, with $K$ = 52 and $K$ = 242 subcarriers, respectively.

\subsection{Deep learning models}
\label{sec:NNmodels}
To test the proposed method and quantify the improvement over the bibliographic LT method described in section \ref{sec:concepts}, the datasets are manipulated in two different ways. The datasets Dataset A, Dataset B, EHUCount, and OPERAnet are evaluated by applying, on the one hand, a fully-connected neural network (FNN) and, on the other hand, a convolutional neural network (CNN). Stratified shuffle split cross-validation \cite{ssscv} with 5 iterations are used in the training of both networks.

In turn, the ReWiS dataset is evaluated using ProtoNet\cite{protonet}, a few shot learning (FSL) strategy \cite{fsl}, as described in his work\cite{rewis}.

\subsubsection{Fully-connected Neural Network} 
\begin{figure}[hbt]
\centering
\includegraphics[width=0.55\textwidth]{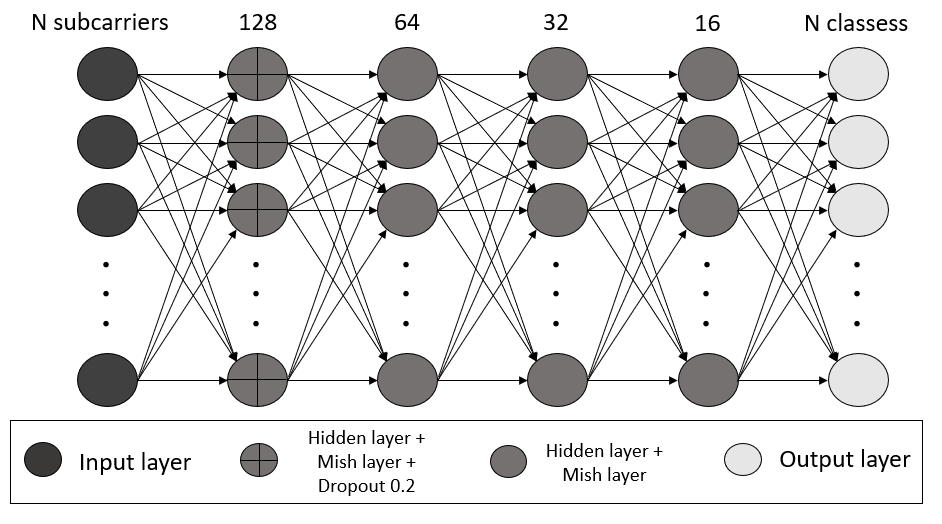}
\caption{Fully-connected neural network as used in this work for WiFi networks.}
\label{dense_image}
\end{figure}

In this case, CSI phase data is classified individually per OFDM symbol, assigning each one the label that corresponds to it. This way, if the dataset has $K$x$S$ dimensions, $1$x$S$ labels are assigned. The datasets are evaluated using a full-connected neural network with four hidden layers. In addition, Mish activation layers \cite{mish} are introduced between the hidden layers to improve the information transmitted by the network using one of the new functions layer developed. Finally, a dropout layer of coefficient 0.2 is placed after the first hidden layer to avoid overfitting. The number of neurons of the first, second, third, and fourth hidden layers is 128,  64, 32, and 16, respectively, for WiFi CSI. DVB-T2 numbers are 256, 128, 64, and 32. The last layer has the same neurons as the classes to be classified. An example of this FNN is depicted in Fig. \ref{dense_image}.

\subsubsection{Convolutional Neural Network} To consider a sufficient time interval in which environmental changes may occur, CSI data are grouped into clusters and evaluated using a CNN. In each dataset, these groups are formed by a different number of symbols. The resizing of the data in the input network to make square inputs that can be used as images must consider the number of subcarriers, which changes for each dataset. Therefore, the dimensions of the inputs of this network are ($r$, $r$, 2), where $r$ is equal to 128 or 256 in WiFi datasets or Dataset A, respectively, and 2 is due to that phase and amplitude are used. This way, the input obtained is comparable to a two-color square image.

\begin{figure*}[hbt]
\centering
\includegraphics[width=1\textwidth]{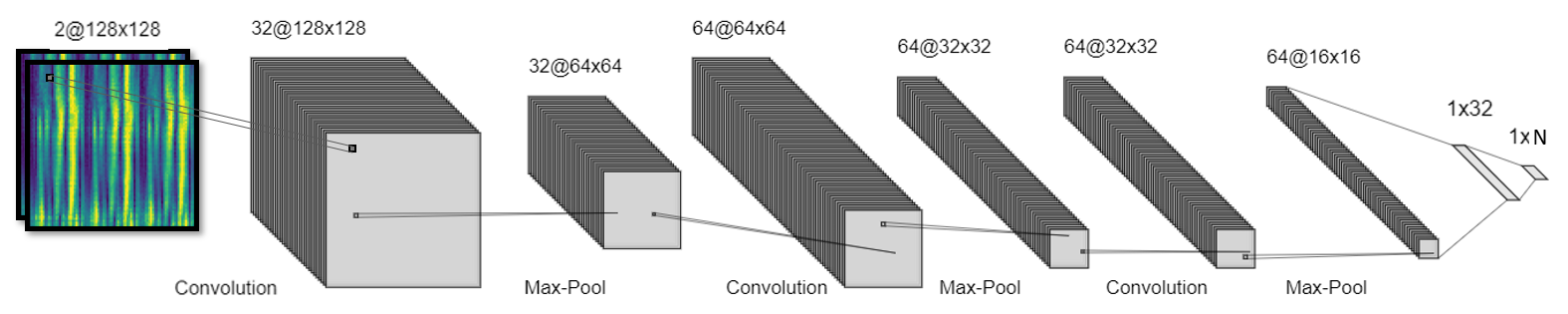}
\centering
\caption{Example of a Convolutional Neural Network as used in this work, with an input of size (128,128,2) and output of size N classes.}
\label{CNN_image}
\end{figure*}

This network consists of a two-channel input layer of size ($r$, $r$, 2) and three two-dimensional convolutional layers with 64, 32, and 32 neurons with three max-pooling layers between them. Behind the convolutional layers is a flattened layer to vectorize the output. Then, there are two full-connected layers, one with 32 neurons and the last one with the number of classification classes. An example of this CNN is depicted in Fig. \ref{CNN_image}.

\subsubsection{Few Shot Learning} 
In the case of the ReWiS dataset, the objective is to replicate the processing performed by its authors at \cite{rewis}, so FSL  ProtoNet processing is applied to the raw data. The goal of FSL is to generalize quickly to new tasks containing only a few samples with supervised information. ProtoNet is based on the idea that there is an environment in which points are clustered around a single prototypical representation for each class.

In this case, four activities are classified: empty, walk, stand, and jump. First, each activity set is divided into intervals of 300 symbols. Single Value Decomposition is applied to each interval to reduce its dimension from $SxK$ to $KxK$. Finally, the linear correlation coefficient, or Pearson's coefficient, is applied to this matrix, obtaining another matrix of linear coefficients, $KxK$, which is the training network's input. The training is performed by means of a CNN with four convolutional blocks. Each block comprises a 64-filter 3×3 convolution, a batch normalization layer, a ReLU nonlinearity, and a 2×2 max-pooling layer that is applied after each of the blocks. 

\section{Results and discussion}
\label{sec:results}

In this section, several comparative analyses are carried out using the aforementioned datasets and DL models. To do that, different classification problems of HAR (people counting, position detection, and gesture recognition) are solved through the CSI information, i.e., using CSI amplitude, CSI phase, or combining amplitude and phase. Furthermore, six different phase processing methods (LT, LRT, LRT + SG filtering in the frequency domain, LRT + SG filtering in the time domain, LRT + two-dimensional SG filtering, and the TSFR proposal) are compared when CSI phases feed the described DL models. Accuracies given by the raw values of the CSI are considered as a benchmark of the models. Performance results are given in terms of averaged percentage values of accuracy along with the standard deviation. The average and the standard deviation are computed when several scenarios or receivers are provided in the same dataset.

Tables \ref{FCNN_count} and \ref{FCNN_pos} show the accuracy results for people counting and position detection, respectively, using the FNN model. In this network, the values of amplitude and phase, as well as the combination of both, are used separately. The amplitude accuracy is constant in all columns of Tables \ref{FCNN_count} and \ref{FCNN_pos} as it does not change after phase processing. Firstly, Table \ref{FCNN_count} shows that using amplitude versus raw phase gives better values for the Dataset A and OPERAnet datasets. In contrast, the results for Dataset B and EHUCount are similar. It is also noted that the LRT calibration of the phase gives better results than the LT method for two of the four datasets, while the accuracy is the same for the other two. Regarding smoothing, the SG filter gives better results for the time than the frequency in three of the four datasets analyzed. Still, in OPERAnet, the time smoothing generates a very low accuracy. On the other hand, 2D smoothing gives better results than frequency or time in all cases. However, the TSFR method using only the phase is the one that offers the best results of all, maintaining in all cases accuracies above 94\%. Table \ref{FCNN_pos} shows the results for classifying fixed positions in Datasets A and B. While the analysis is similar to that in Table \ref{FCNN_count}, the accuracy of the TSFR method on Dataset B is striking, as it achieves 99\% accuracy compared to 33\% for time smoothing, in both cases working only with the phase. 

In Tables \ref{CNN_count} and \ref{CNN_pos}, the performance of counting people and detecting position, respectively, is given for the proposed CNN model. In this case, the components of the CSI have been windowed to create images as inputs of the model. This network combines amplitude and phase values in matrices of the form ($r$, $r$, 2), as explained in Section \ref{sec:NNmodels}. In these tables, we can see performance indicators similar to those in Tables \ref{FCNN_count} and \ref{FCNN_pos}. First, the LRT method still offers better or equal accuracies than the LT method.

\begin{table*}[h!]
\caption{Accuracy values, in \%, of all datasets using a Fully-connected Neural Network and classifying by number of people in the room.}
\centering
\setlength{\tabcolsep}{3pt}
\begin{tabular}{|l|l|l|l|l|l|l|l|l|}
\hline
Dataset&
Variable&
raw  &     
LT &  
LRT &     
LRT  + SG freq &    
LRT  + SG time &  
LRT  + SG 2D &
TSFR\\
\hline
\multirow{3}{*}{Dataset A} & abs &  74$\pm$5 &  74$\pm$5 & 74$\pm$5 &  74$\pm$5 &  74$\pm$5 & 74$\pm$5 & 74$\pm$5\\
& phase &  65$\pm$4 & 81$\pm$5 & 86$\pm$1 & 85$\pm$3 & 96$\pm$1 & 95$\pm$1 & \textbf{97$\pm$1}\\
& abs+phase &  72$\pm$8 &  88$\pm$3 &   87$\pm$1 &   89$\pm$1 &  93$\pm$5 &   96$\pm$1 & 94$\pm$1\\
\hline
\multirow{3}{*}{Dataset B} & abs & 98.1 & 98.1 & 98.1 & 98.1 & 98.1 & 98.1 & 98.1\\
& phase & 98.2 & 98.3 & 98.5 & 98.3 & 98.7 & 98.7 & \textbf{99.9}\\
& abs+phase & 98.4 & 98.3 & 98.2 & 98.1 & 98.7 & 98.3 & 99.3\\
\hline
\multirow{3}{*}{EHUCount} & abs & 80$\pm$7 & 80$\pm$7 & 80$\pm$7 & 80$\pm$7 & 80$\pm$7 & 80$\pm$7 & 80$\pm$7\\
& phase &  79$\pm$9 &  79$\pm$9 &  86$\pm$5 & 87$\pm$5 &  \textbf{99.7$\pm$0.1} & \textbf{99.8$\pm$0.1} & \textbf{99.7$\pm$0.1}\\
& abs+phase & 84$\pm$7 & 85$\pm$7 & 88$\pm$4 & 89$\pm$2 &  \textbf{99.8$\pm$0.2} & 98.8$\pm$0.8 & \textbf{99.6$\pm$0.2}\\
\hline
\multirow{3}{*}{OPERAnet} & abs &  82$\pm$5 &  82$\pm$5 & 82$\pm$5 &  82$\pm$5 &  82$\pm$5 & 82$\pm$5 & 82$\pm$5\\
& phase &  50$\pm$2 &  54$\pm$5 &   54$\pm$4 &   81$\pm$3 &  30$\pm$1 &   88$\pm$1 & \textbf{94$\pm$1}\\
& abs+phase &  65$\pm$1 &  65$\pm$1 &   65$\pm$1 &   84$\pm$1 &  64$\pm$1 &   85$\pm$1 & 84$\pm$1\\
\hline
\end{tabular}
\label{FCNN_count}
\end{table*}

\begin{table*}[h!]
\caption{Accuracy values, in \%, of Datasets A and B using a Fully-connected Neural Network and classifying by fixed position of people in the room.}
\centering
\setlength{\tabcolsep}{3pt}
\begin{tabular}{|l|l|l|l|l|l|l|l|l|}
\hline
Dataset&
Variable&
raw  &     
LT &  
LRT &     
LRT + SG freq &    
LRT + SG time &  
LRT + SG 2D &
TSFR\\
\hline
\multirow{3}{*}{Dataset A} & abs &  80$\pm$3 &  80$\pm$3 & 80$\pm$3 &  80$\pm$3 &  80$\pm$3 & 80$\pm$3 & 80$\pm$3\\
& phase &  61$\pm$1 & 84$\pm$1 & 86$\pm$2 & 86$\pm$2 & 95$\pm$1 & \textbf{97$\pm$1} & \textbf{96$\pm$1}\\
& abs+phase &  75$\pm$5 &  86$\pm$3 &   83$\pm$4 &   90$\pm$1 &  90$\pm$3 &   91$\pm$2 & 91$\pm$2\\
\hline
\multirow{3}{*}{Dataset B} & abs & 24 & 24 & 24 & 24 & 24 & 24 & 24\\
& phase & 24 & 24 & 26 & 25 & 33 & 26 & \textbf{99}\\
& abs+phase & 24 & 23 & 24 & 26 & 30 & 25 & 91\\
\hline
\end{tabular}
\label{FCNN_pos}
\end{table*}
\begin{table*}[h!]
\caption{Accuracy values, in \%, of all datasets using a Convolutional Neural Network and classifying by number of people in the room.} 
\centering
\setlength{\tabcolsep}{3pt}
\begin{tabular}{|l|l|l|l|l|l|l|l|l|}
%{|p{40pt}|p{30pt}|p{30pt}|p{30pt}|p{35pt}|p{35pt}|p{35pt}|p{35pt}|}
\hline
Dataset&
raw  &     
LT &  
LRT &     
LRT + SG freq &    
LRT + SG time &  
LRT + SG 2D &
TSFR\\
\hline
Dataset A &  46$\pm$8 &  78$\pm$5 &   81$\pm$7 &   82$\pm$9 &  53$\pm$5 &   80$\pm$7 & \textbf{91$\pm$3}\\
\hline
Dataset B & 98.4 & 98.6 & 98.3 & 97.9 & 98.7 & 98.3 & \textbf{99.9}\\
\hline
EHUCount & 34$\pm$6 & 55$\pm$7 & 58$\pm$4 & 75$\pm$9 &  90$\pm$4 & 94$\pm$2 & \textbf{96$\pm$2}\\
\hline
OPERAnet  &  56$\pm$3 &  55$\pm$2 &   54$\pm$4 &   55$\pm$3 &  77$\pm$1 &   90$\pm$3 & \textbf{99.9$\pm$0.1}\\
\hline
\end{tabular}
\label{CNN_count}
\end{table*}

\begin{table*}[h!]
\caption{Accuracy values, in \%, of Datasets A and B using a Convolutional Neural Network and classifying by fixed position of people in the room.}
\centering
\setlength{\tabcolsep}{3pt}
\begin{tabular}{|l|l|l|l|l|l|l|l|l|}

\hline
Dataset&
raw  &     
LT &  
LRT &     
LRT + SG freq &    
LRT + SG time &  
LRT + SG 2D &
TSFR\\
\hline
Dataset A & 29$\pm$22 &  62$\pm$17 &   60$\pm$8 &   68$\pm$15 &  42$\pm$9 &   56$\pm$6 & \textbf{80$\pm$8}\\
\hline
Dataset B & 24 & 25 & 26 & 25 & 32 & 26 & \textbf{96}\\
\hline
\end{tabular}
\label{CNN_pos}
\end{table*}

\begin{table*}[h!]
\caption{This table is an extension of Tables \ref{FCNN_count} and \ref{CNN_count} for Dataset B. Since this is an unbalanced dataset, accuracy results are shown for each class (0, 1, and 2 people).}
\centering
\setlength{\tabcolsep}{3pt}
\begin{tabular}{|p{40pt}|p{40pt}|p{10pt}|p{10pt}|p{10pt}|p{10pt}|p{10pt}|p{10pt}|p{10pt}|p{10pt}|p{10pt}|p{13pt}|p{13pt}|p{13pt}|p{12pt}|p{12pt}|p{12pt}|p{10pt}|p{10pt}|p{10pt}|p{10pt}|p{10pt}|p{12pt}|p{13pt}|p{12pt}|}
\hline
&
&
\multicolumn{3}{c|}{raw} &     
\multicolumn{3}{c|}{LT} &  
\multicolumn{3}{c|}{LRT} &     
\multicolumn{3}{c|}{LRT + SG freq} &    
\multicolumn{3}{c|}{LRT + SG time} &  
\multicolumn{3}{c|}{LRT + SG 2D} &
\multicolumn{3}{c|}{TSFR}\\
\hline
Network & Variable &\textit{0} &\textit{1} &\textit{2} &\textit{0} &\textit{1} &\textit{2}&\textit{0} &\textit{1} &\textit{2}&\textit{0} &\textit{1} &\textit{2}&\textit{0} &\textit{1} &\textit{2}&\textit{0} &\textit{1} &\textit{2}&\textit{0} &\textit{1} &\textit{2}\\
\hline
\multirow{3}{*}{FNN} & abs & 0 & 98 & 98 & 0 & 99 & 98 & 0& 98 & 99 & 0 & 98 & 98 & 0 & 98 & 98 & 0 & 98 & 98 & 0 & 98 & 98\\
& phase & 0 & 98 & 98 & 0 & 98 & 98 & 0& 98 & 98 & 0 & 99 & 98 & 0 & 98 & 98 & 0 & 98 & 99 & \textbf{99} & \textbf{100} & \textbf{100}\\
& abs+phase & 0 & 98 & 98 & 0 & 98 & 98 & 0& 98 & 99 & 0 & 98 & 98 & 0 & 99 & 98 & 0 & 98 & 98 & 99 & 99 & 100\\
\hline
\multirow{1}{*}{CNN} & abs+phase & 0 & 98 & 99 & 0 & 98 & 99 & 0& 98 & 99 & 0 & 98 & 99 & 0 & 99 & 98 & 0 & 99 & 98 & \textbf{99} & \textbf{100} & \textbf{100}\\
\hline
\end{tabular}
\label{scen2}
\end{table*}

 Comparing time smoothing with frequency smoothing shows that one offers better results in two datasets and the other in the other two, with substantial differences. Also, in this case, 2D smoothing improves the frequency or time smoothing results in all cases, but it is outperformed by the TSFR method, which achieves excellent accuracy of more than 90\% in all cases. 

 Table \ref{CNN_pos} shows the most different results between methods concerning the other tables. For example, frequency smoothing for Dataset A gives better results than time smoothing or 2D smoothing, while for Dataset B, time smoothing is the best.
For example, for Dataset A, frequency smoothing gives better results than time smoothing or 2D smoothing, while for Dataset B, time smoothing is the best. Moreover, the accuracy values in both datasets are the lowest of all. Nevertheless, the TSFR method remains the best, with reasonable accuracy values. The accuracy of the TSFR method in Dataset B is 96\%, compared to the second highest, 33\%, and in Dataset A is 88\%, while the second highest is 68\%.

In addition, we have seen fit to include Table \ref{scen2} to show the Dataset B metrics in more detail, as the high accuracy in Tables \ref{FCNN_count}, and \ref{CNN_count} can be misleading. Dataset B is an unbalanced dataset in which class \textit{0} occupies 1\% of the total size, while classes \textit{1} and \textit{2} are 50\% and 49\%, respectively. Table \ref{scen2} is an extension of Tables \ref{FCNN_count} and \ref{CNN_count}. It shows that, although the overall accuracy values are 98\%, the only method capable of correctly classifying the unbalanced class is the TSFR method.

\begin{figure}[hbt]
\centering
\begin{subfigure}[b]{0.23\textwidth}
    \centering
    \includegraphics[width=\textwidth]{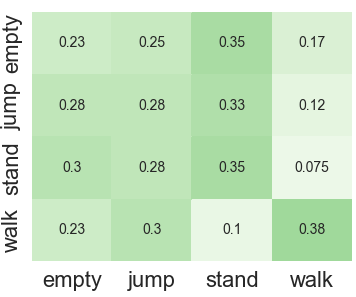}
    \caption{ReWis method, 20MHz BW, Acc=0.597}
\end{subfigure}
\begin{subfigure}[b]{0.23\textwidth}
    \centering
    \includegraphics[width=\textwidth]{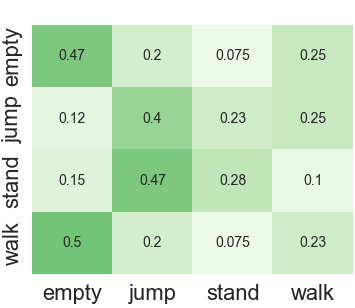}
    \caption{ReWis method, 80MHz BW, Acc=0.782}
\end{subfigure}
\begin{subfigure}[b]{0.23\textwidth}
    \centering
    \includegraphics[width=\textwidth]{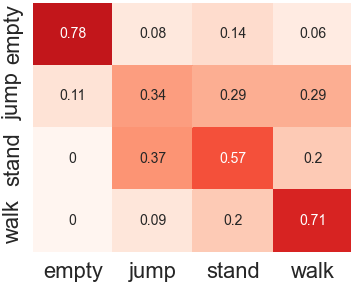}
    \caption{Raw phase, 20MHz BW, Acc=0.306}
\end{subfigure}
\begin{subfigure}[b]{0.23\textwidth}
    \centering
    \includegraphics[width=\textwidth]{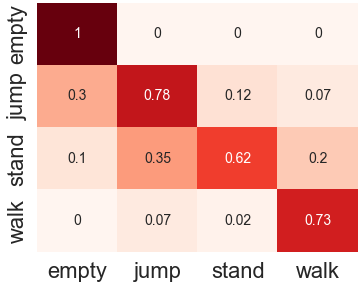}
    \caption{Raw phase, 80MHz BW, Acc=0.344}
\end{subfigure}
\begin{subfigure}[b]{0.23\textwidth}
    \centering
    \includegraphics[width=\textwidth]{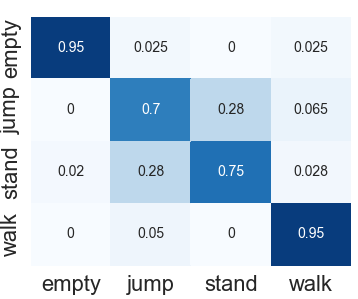}
    \caption{TSFR phase, 20MHz BW, Acc=0.825}
\end{subfigure}
\begin{subfigure}[b]{0.23\textwidth}
    \centering
    \includegraphics[width=\textwidth]{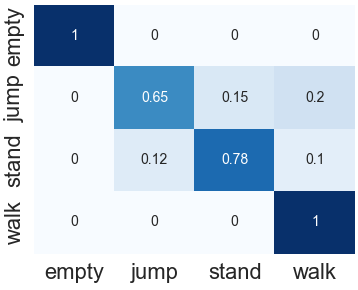}
    \caption{TSFR phase, 80MHz BW, Acc=0.857}
\end{subfigure}
\caption{Confusion matrices: \textbf{(a)} and \textbf{(b)} are based on the ReWis method \cite{rewis} using CSI amplitude. \textbf{(c)} and \textbf{(d)} are achieved with the raw CSI phase. \textbf{(e)} and \textbf{(f)} are based on the TSFR method. The data corresponds to the configuration of single antennas in transmission and reception.}
\label{fig:rewis}
\end{figure}

The ReWiS dataset is analyzed using FSL under the ProtoNet model. In this case, the amplitude and phase values obtained are compared using, on the one hand, the raw CSI values and, on the other hand, the CSI values processed with the TSFR method. The Fig. \ref{fig:rewis} shows the confusion matrix for each comparison for 20 and 80 MHz, including reference values from \cite{rewis}. TSFR phase processing improves the CSI raw results from 32\% to 82\% and from 35\% to 85\% at 20 and 80MHz of bandwidth, respectively, for phase accuracy. These results outperform the accuracy using the amplitude. Testing the TSFR method on this dataset using FSL implies that the method supports the extraction of certain features on the processed phase and improves the transferability of its results between different scenarios.

To summarize, the results indicate that the TSFR method can improve the classification accuracy by counting people, determining their fixed position, and detecting activities using regular neural networks, as shown in all datasets. The success of the Eq.\eqref{phifinal} is observed by comparing the results of the SG filter in the time domain (LRT+SG time) vs. the TSFR method. In all cases, the proposed method in Eq.\eqref{phifinal} to rebuild the distortions generated by the SG filtering in the frequency domain substantially improves the classification algorithms. The accuracy of the TSFR method is always higher than that given by the exclusive application of the SG filter in any domain, including both simultaneously. Moreover, TSFR obtains good results when DL algorithms use CSI data directly, and it also improves results when feature engineering is carried out, as we have observed in the comparative analysis based on the FSL model. Furthermore, it is worth mentioning that the results seem to indicate that their use in unbalanced datasets may help to improve the accuracy in detecting under-represented classes.

\section{Conclusion \& future work}
\label{sec:conclusions}
This paper presents a new method of CSI phase processing for human activity recognition in wireless sensing. The proposed method, coined Time Smoothing and Frequency Rebuild (TSFR), corrects the linear errors of the phase using a linear regression transformation for each CSI. After that, phase smoothing in the time domain is carried out using the SG filter, and finally, filtering distortions are rebuilt in the frequency domain using information from the sanitized phase.
The method presented in this paper serves as a generalization of the method for five different datasets, each of one with a different number of subcarriers and measured with different hardware. In this sense, in future work, some of the parameters used in this study may be studied in more depth for specific datasets. On the one hand, the values chosen for the Savitzky-Golay filter, and on the other hand, the value that multiplies sigma in the Gaussian filter, Eq.\eqref{ds}, which determines the number of data considered outliers. Each dataset may allow a more adjusted value depending on the number of subcarriers, symbols, data quality, and other specifications in both cases.

In conclusion, the comprehensive analysis of the proposed method shows that the TSFR solution outperforms other solutions of phase processing in five CSI datasets provided by different OFDM-based wireless systems, variable configurations, and different HAR activities (people counting (sitting or walking), position detection, and activity recognition). In addition, three different DL networks have been employed. Therefore, the TSFR method allows the use of the CSI phase as a robust source of information for HAR in different conditions of wireless sensing.

\section*{Data availability}
This work contains data extracted from five different datasets based on HAR. The data from OPERAnet, EHUCount, and ReWiS are public and available online. The other two datasets have been developed for this work and are available under request to the corresponding author.

\bibliography{library}

\section*{Funding}

This work has been financially supported by the Basque Government (under grant IT1436-22) and by the Spanish Government (under grant PID2021-124706OB-I00, funded by MCIN/AEI/10.13039/501100011033 and by ERDF A way of making Europe).

\section*{Author contributions}

Investigation, G.D.; methodology, I.S and I.E; resources, J.C.; software, G.D.; supervision, M.V.; writing—original draft, G.D, I.S. and I.E.; writing—review and editing, I.L. and M.V. Dataset A measurement, G.D, I.S, I.E, I.L. and M.V.; Dataset B measurement, J.C. All authors have read and agreed to the published version of the manuscript.

\section*{Competing Interests }
The authors declare no competing interests.

\end{document}